\documentclass[12pt,preprint]{aastex}

\usepackage{CJK}

\usepackage{epsfig}

\shorttitle{Molecules in 21 and 30\,$\mu$m sources}

\shortauthors{Zhang}

\begin{document}

\title{Molecular gas in 21 and 30 micron sources: the 2\,mm and 1.3\,mm spectra of IRAS\,21318+5631 and 22272+5435}

%\begin{CJK*}{Bg5}{bsmi}
%\CJKtilde
\begin{CJK*}{UTF8}{gbsn}

\author{Yong Zhang}

\affil{School of Physics and Astronomy, Sun Yat-sen University, 2 Daxue Road, Tangjia, Zhuhai,
Guangdong Province,  China}
%\affil{ Laboratory for Space Research,
%Faculty of Science, The University of Hong Kong, Pokfulam Road, Hong Kong, China}
\email{zhangyong5@mail.sysu.edu.cn}

\begin{abstract} 

The carriers of the 21 and 30\,$\mu$m emission features  in infrared spectra of circumstellar envelopes are a long-standing enigma. In this paper, we present the results of molecular line observations toward two circumstellar envelopes exhibiting the 21 and/or 30\,$\mu$m features, 
IRAS\,21318+5631 and 22272+5435, aiming at investigating whether they have unusual gas-phase chemistry and searching for possible gas-phase precursor of the carriers of the two dust features.
The spectra cover several discrete frequency ranges of 130--164\,GHz and  216.5--273\,GHz, resulting in a detection of 13 molecular species and isotopologues in each object. Rotation-diagram analysis is carried out to determine the molecular abundances, column densities, and excitation temperatures. We did not discover any molecular species that is unexpected in a normal C-rich star. Nevertheless, there exists subtle difference between their molecular abundances. IRAS\,22272+5435 shows stronger SiC$_2$ and HC$_3$N lines and weaker SiS lines than IRAS\,21318+5631, presumably suggesting that this 21\,$\mu$m source is more carbon rich and has experienced a more efficient dust formation. We discuss the potential implications of the results for the carriers of the 21\,$\mu$m and 30\,$\mu$m features.

\end{abstract}

\keywords{
ISM: molecules --- circumstellar matter --- line: identification ---
stars: AGB and post-AGB}

\maketitle
\end{CJK*}

\section{Introduction}

Circumstellar envelopes (CSEs) surrounding low- and intermediate-mass evolved stars are a major engine driving the material cycle in the Galaxy.
The evolution of CSEs can be divided into three stages: asymptotic giant branch (AGB),
proto-planetary nebula (PPN), and planetary nebula (PN), during which
atoms and molecules ejected from stellar surface are processed into larger molecules and dust, and finally disperse into the ISM. With the advance of observational technique,
over 70 different gas-phase molecules have been unambiguously detected in CSEs so far
\citep[][]{cer11,mc18}.
However, the whole picture of circumstellar chemistry during the AGB-PPN-PN
evolution is still far from complete.
A challenging issue in this area is to identify some infrared emission features,
such as those at 21 and 30\,$\mu$m. The two features may
contribute to up to 10 and 30 percent of the total infrared luminosity of
CSEs \citep{vx00,ml15}, respectively. Their carriers, no doubt, contain essential information on chemical processes in circumstellar environments.

The 21\,$\mu$m band, first discovered by \citet{kv89}, is a rare feature that has been detected in less than three dozens of C-rich PPNs
\citep{hv09,vol10,ml16, gm19}. This feature
in different sources exhibits an identical peak wavelength
($\sim20.1$\,$\mu$m) and a remarkably similar profile \citep{vol99}.
Its central wavelength was measured to be $20.47\pm0.10$\,$\mu$m \citep{sloan}.
A variety of candidates have been proposed as its carrier, such as SiS$_2$, FeO, TiC, hydrogenated fullerenes, 
polycyclic aromatic hydrocarbons, hydrogenated amorphous carbon, and nanodiamonds 
\citep[see, e.g.,][and references therein]{spc04,pos04},
but no consensus has been reached. Some carrier candidates were examined by \citet{zha09c}, who
found that all of them except FeO are not abundant enough  to account for the strong 21\,$\mu$m 
emission, or would produce some sub-features which have never been found in 21\,$\mu$m sources. 
Subsequently, \citet{km17} showed that FeO particles are sensitively
affected by oxygen fugacity and temperature, and its laboratory spectra vary depending on
defects and disorders, suggesting that FeO is unlikely to be the carrier of the 21\,$\mu$m 
feature.  As PPNs represent a very short-lived phase between the 
AGB and PN stages, the carrier of this feature  is presumably associated with a rapid chemical process, or alternatively, is formed in the AGB phase but not excited until the post-AGB phase \citep{ml16}.

The 30\,$\mu$m feature is more commonly detected than the 21\,$\mu$m one.  It has been discovered in C-rich AGB stars, PPNs, and PNs \citep{for81,vol02}.  Although  all the 21\,$\mu$m emitters also exhibit the 30\,$\mu$m feature, there is no correlation between their strengths \citep{om95, ml15}. A statistic study shows that the strengths of the 30\,$\mu$m feature correlate with the dust temperature and the metallicity of host galaxy \citep{gm19}.
Though several candidate carriers have been discussed \citep{nu85}, 
solid magnesium sulfide (MgS) is the mostly widely accepted one
 \citep{goe85}. Large variations
 in the profiles of the 30\,$\mu$m feature have been revealed in 
different sources by \citet{hon02}, who suggested that this is due to
the changes of MgS dust temperature. However, based on energetic and
abundance arguments, \citet{zha09d} show that MgS dust is unlikely
to be responsible for the observed feature strength. Their results are
supported by a detailed study of the shape and position of
the 30\,$\mu$m feature \citep{ms13}.

As the 21 and 30\,$\mu$m features are detected only in C-rich environments, their
carriers are likely to be associated with carbon chemistry. 
Presumably, the routes of gas-phase chemistry and dust formation greatly rely on
the physical conditions and evolutionary stages of CSEs. 
To study gas-phase molecules in the 21 and 30\,$\mu$m sources might provide
useful clues for the accurate identification of the infrared features.
Previous radio observations of these sources mainly focus on strong lines
such as CO lines \citep[e.g.][]{wk90,nk12}.
In this paper, we report new radio spectra of the gas-phase molecules towards two 
CSEs IRAS\,21318+5631 and 22272+5435 covering a wide frequency range.
These data allow us to investigate
the chemical processes in the environments favoring the origin of the two features.
IRAS\,21318+5631 is an extreme carbon star that reaches the optically
obscured superwind phase and is about to leave the AGB stage.
 \citet{kv99} discovered the 
30\,$\mu$m feature in this object. They also reported the detection of an extremely
 weak feature resembling the 21\,$\mu$m feature. This, however, was not confirmed in further measurements 
\citep{cm05}.  We therefore do not consider IRAS\,21318+5631 as a 21\,$\mu$m source.
IRAS\,22272+5435 is a PPN exhibiting relatively strong 21 and 30\,$\mu$m features 
\citep{so97}.
Extremely high C/O ratio ($\sim12$) has been obtained from its optical 
spectrum \citep{zk95}. The difference of chemical compositions in
IRAS\,21318+5631 and 22272+5435 might reflect the chemical processes
in the short transition phase from the AGB to PPN.

As one of the series study of our group on molecular lines in CSEs 
\citep{he08,zy08,zy09a,zy09b,cz12,zy13,zy20}, this paper mainly attempts
to investigate the relationship between gas-phase chemistry and 
the unusual infrared behaviors in post-AGB phase.
These studies, combined with the
molecular observations of evolved stars by other authors 
\citep[e.g.][and the reference therein]{cernicharo00,pardo07,td10,ec14},
can provide a more complete picture of circumstellar chemistry during the AGB-PPN-PN 
transitions.  The remaining parts of the paper are arranged as follows. In Section 2 the 
2\,mm and 1.3\,mm radio spectra towards IRAS\,21318+5631 and 22272+5435
are presented. The detected molecules and derived abundances
are shown in Section 3. Section 4 contains a comparison of the molecular
compositions in different sources and the chemical implications.
 The main conclusions are summarized in Section 5.

\section{Observations and data reduction}

The observations at the 2\,mm and 1.3\,mm windows were carried out in 2009 January and 
2014 April with the {\it Arizona Radio Observatory (ARO)} 12\,m telescope at Kitt Peak and the {\it Heinrich Hertz Submillimeter Telescope (SMT)} 10\,m telescope at Mt. Graham.  
The beam switching mode was used with an azimuth beam throw of {2\arcmin}.
Pointing and focus were checked every 1--2 hours using nearby planets, and typically
the pointing offsets were less than 10$\arcsec$. 
The dual-channel SIS receivers were employed, operated in single sideband dual 
polarization mode. The image rejection ratio is typically better than $18$\,dB.
The system noise temperatures were typically 250\,K and 400\,K for the 2\,mm and 
1.3\,mm bands, respectively.  The integration times per tuning step were 2--4 hours.
The half power beam widths ($\theta_b$) of the 12m telescope and the SMT are about 40$''$  
and 30$''$ at the observed frequency ranges, respectively.
For the 2\,mm observations, the spectrometer backends 
employed are two 256-channel filter banks (FBs) with a channel width of 1\,MHz 
and a  millimeter autocorrelator (MAC) with 3072 channels and 195\,kHz resolution.
For the 1.3\,mm observations, the 2048-channel acousto-optical spectrometer (AOS)
and  the 1024-channel Forbes Filterbanks (FFBs) were used simultaneously with
a channel width of 500\,kHz and 1\,MHz, respectively. The temperature scales
of the 2\,mm and 1.3\,mm are given in term of $T^*_{\rm R}$ and  $T^*_A$, respectively.
The main beam temperature $T_{\rm R}$  then was derived by $T^*_{\rm R}/\eta^*_m$ and $T^*_A/\eta_{mb}$, where the corrected beam efficiency  $\eta^*_m=0.8$ and 
the main beam efficiency $\eta_{mb}=0.7$.

The observed frequency range is from 130--164\,GHz and 216.5--273\,GHz. Our original 
proposal is to perform an unbiased line surveys at the two windows. 
However, because of 
limited observation time and occasional poor weather conditions,  we are not able to
observe the whole frequency range and there remain several gaps in the observed windows. 
The priority of our observations was given to the frequency ranges where strong lines
have been previously observed by us in other sources \citep[e.g. IRC+10216;][]{he08}. 
 The strengths of molecular lines in the two IRAS sources are intrinsically much fainter than 
those in IRC+10216. The CO ($J=2$--1) line strengths in the two sources 
are lower than that in IRC+10216 by two orders of magnitude. If assuming
that all the line strengths follow the same scale, we speculate that those lying within 
the unobserved spectral gaps should be below our detection sensitivity.

The data reductions were  done using the CLASS software package in GILDAS\footnote{GILDAS is developed and distributed by the Observatorie de Grenoble and IRAM.}.
 following the
standard procedure. Discarding the scans due to bad atmospheric conditions and receiver 
instabilities, we co-added
 the calibrated spectral data using the $rms$ noise of each spectrum as weights.
The spectral baselines were fit by low-order polynomials to the line-free spectral regions.
The 2\,mm and 1.3\,mm spectra were smoothed and rebinned to a frequency resolution of 
1\,MHz and 3\,MHz, resulting in a typical $rms$ noise levels of $\sim5$\,mK in main beam 
temperature unit. 

\section{Results}

\subsection{The overall observations}

\begin{figure*}
\epsfig{file=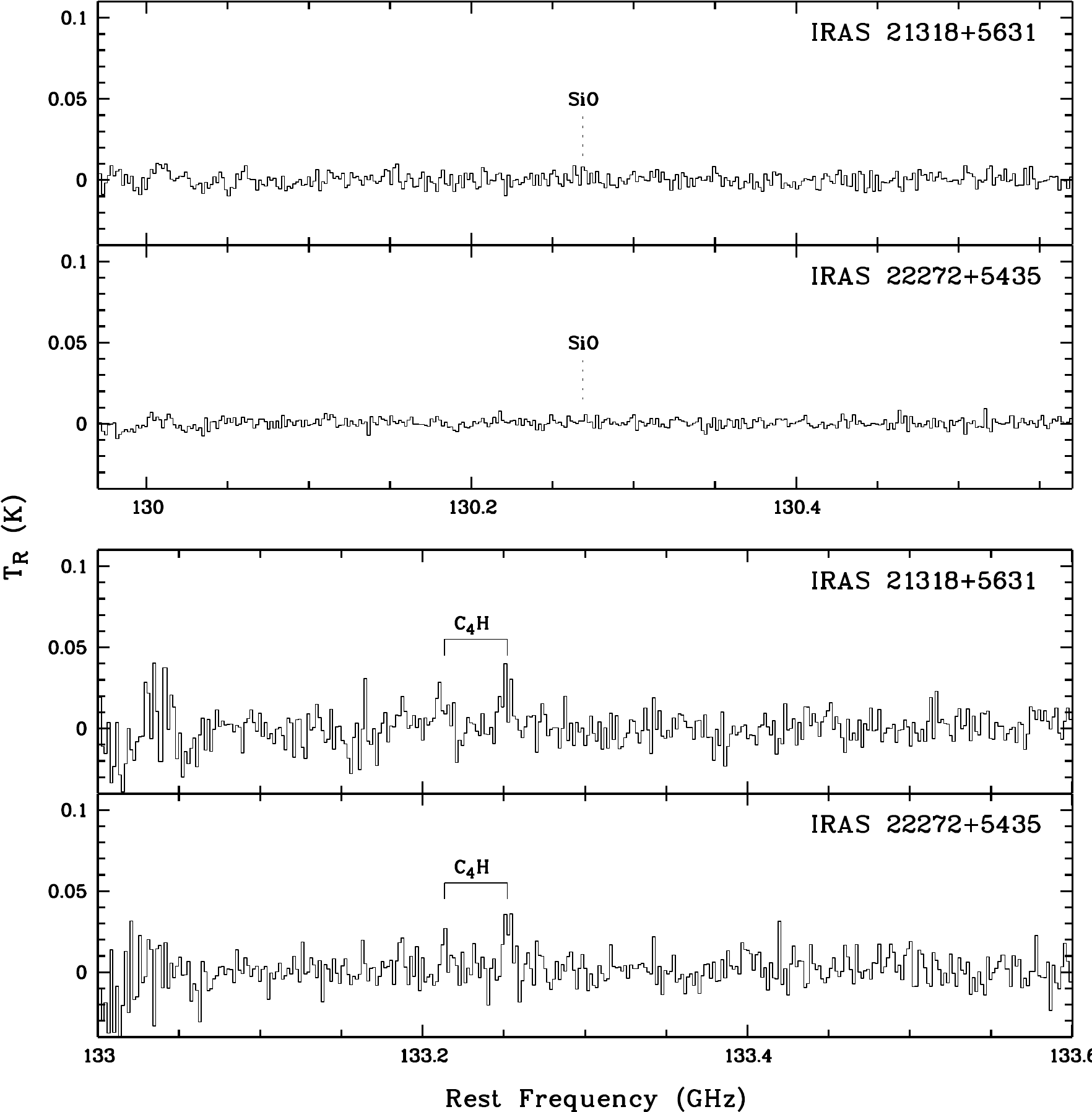}
\caption{The 12m spectra of IRAS\,$21318+5631$ and $22272+5435$. The dotted vertical lines mark the undetected 
or marginally detected transitions that are strong in IRC+10216. Note that there are some artificial features at the edge of each
 band thanks to the bandpass irregularities of the MAC.}\label{f1}
\end{figure*}

\begin{figure*}
\epsfig{file=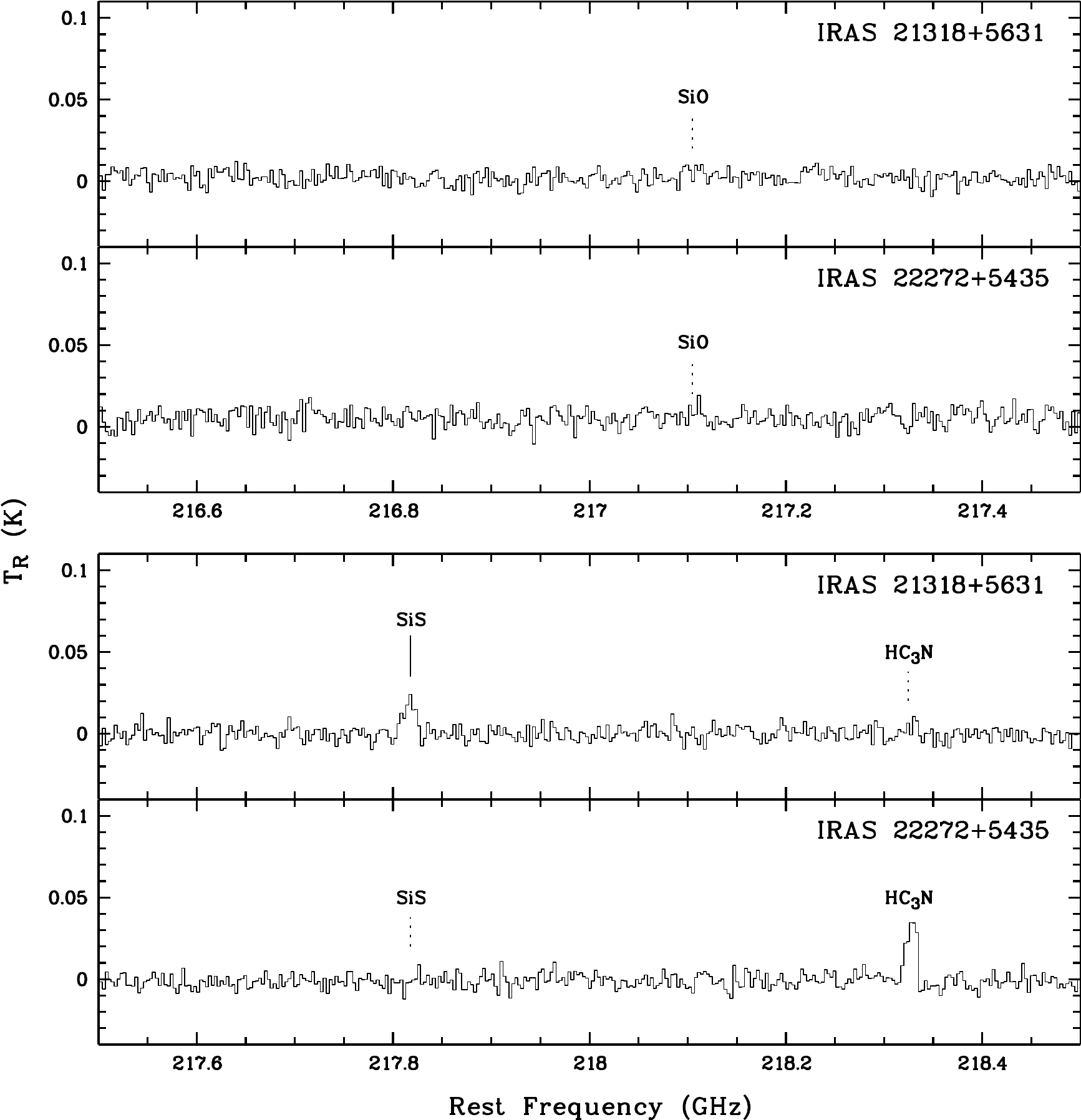,height=16cm}
\caption{The SMT spectra of IRAS\,$21318+5631$ and $22272+5435$.
The dotted vertical lines mark the undetected or marginally detected
transitions that are strong in IRC+10216.
%Note that the CO and $^{13}$CO lines in IRAS\,$21318+5631$  show foreground Galactic contamination.
}\label{f2}
\end{figure*}

The obtained 2\,mm and 1.3\,mm spectra, along with line identifications, are presented in Figures~\ref{f1} and \ref{f2}, respectively.  
The features above 3$\sigma$ noise levels are considered as real detections.
Although some extremely weak features have peak strengths lower than 3$\sigma$ 
noise level, they are regarded as positive detections 
if other transitions from the same species are clearly detected. The line 
identifications are carried out using the archives of molecular line frequencies derived from the theoretical calculations of the JPL catalog \citep{pickeet98}\footnote{http://spec.jpl.nasa.gov.} and the Cologne database for molecular spectroscopy \citep[CDMS,][]{muller01,muller05}\footnote{http://www.ph1.uni-koeln.de/vorhersagen/.}. We discover ten molecular species and three isotopologues in each
source.  Positively detected species in both sources include  C$_2$H, C$_4$H, HC$_3$N, SiC$_2$, CS, C$^{34}$S, CO, $^{13}$CO, CN, HCN, H$^{13}$CN, and HNC. For the first time, CH$_3$CN is detected in IRAS\,$22272+5435$, confirming its
extreme C-rich rich nature.

The overall spectral properties of the two sources are consistent with their C-rich
nature. For the first glance, there is no essential difference between their
spectra and that of the prototypical carbon star IRC+10216 obtained by
\citet{he08} and \citet{td10}.  But we did not detect SiO and HC$^{15}$N, two
molecules having strong transitions in the spectrum of IRC+10216 within the same frequency
range. 
Table~\ref{trans} shows a full list of line identifications and measurements,
in which columns~1--3 respectively give the frequencies, identified species, and transitions,
and the remaining columns list the $rms$ noise levels, the peak and integrated intensities,
and the line widths ($FWHM$) obtained by fitting Gaussian line profiles.
For non-detected lines, the upper limits, corresponding to
$3\sigma$ of the peak and integrated intensities, are given.

The most interesting result of our observations is the apparently different relative 
intensities of the HC$_3$N, SiC$_2$, and SiS lines in the two IRAS sources
(see, e.g., the spectral ranges from 163--164, 217.5--218.5, 235.5--237.5, 272--273\,GHz in
Figures~\ref{f1} and \ref{f2}). IRAS\,$22272+5435$  clearly exhibits  a number of lines from HC$_3$N and SiC$_2$, while the two molecules are only marginally detected in IRAS\,$21318+5631$.
Although the line intensities in IRAS\,$22272+5435$ are generally stronger than
those in IRAS\,$21318+5631$, it does not exhibit SiS emission that is clearly detected in the later source.

\subsection{Line profiles}

All the lines in IRAS\,$22272+5435$ show a notably narrow single-peaked profile 
with a $FWHM$ of about 10\,km\,s$^{-1}$, much lower  than the typical value of 
20\,km\,s$^{-1}$ previously obtained for other CSEs by us.
No high-velocity wing can be seen for any line.  This is compatible with the CO observation of \citet{hb05},
probably implying that the carrier of the 21\,$\micron$ feature could be
destructured if there exist high-velocity shocks.  
The low expansion velocity, together with the fact that 21\,$\micron$ sources have
high mass-loss rates \citep{hk91}, indicates a high density, which might favors the 
condensation of gas-phase molecules and the formation
of the carrier of the 21\,$\micron$ feature.

In contrast, the molecular lines in IRAS\,$21318+5631$ consistently
appear to be broader with a $FWHM$ ranging from 15--25\,km/s. C$_4$H and C$_2$H
lines show a double-peaked profile. A narrow dip is clearly superimposed on the
center of the CO and $^{13}$CO lines, which could be attributed to 
the presence of an outer cold cloud.  There is no apparently asymmetrical profile revealed 
in both sources, suggesting that the gaseous envelopes
 may have an approximately symmetrical shape.

\subsection{Rotation diagram analysis}

Standard rotation diagram analysis was  performed to derive
molecular excitation temperatures and column densities. Under
the assumption that all the lines are optically thin and thermalized, and
the level populations are in locate thermodynamic equilibrium (LTE),
the relation between the molecular column density $N$
and the integrated line intensity of the source $\int T_s dv$
can be given by:
\begin{equation}
\ln \frac{N_u}{g_u}=\ln\frac{3k\int T_s dv}{8\pi^3\nu S\mu^2}=
\ln\frac{N}{Q(T_{ex})}-\frac{E_u}{kT_{ex}},
\label{boltzmann}
\end{equation}
where $N_u$, $g_u$, and  $E_u$ is
the population, degeneracy, and excitation energy of the upper level,
$S$ the line strength, $\mu$ the dipole moment,
$\nu$ the line frequency,
 $Q$ the rotation
partition function, and $T_{ex}$ the excitation  temperature.
Assuming that the source brightness distribution and the antenna beam have a Gaussian profile,
the source brightness temperature was obtained by $T_s=T_R(\theta^2_b+\theta^2_s)/\theta^2_s$.
The source diameter $\theta_s$ is difficult to determine and may vary from species to species.
In this study, according to the CO image of IRAS\,$22272+5435$ by \citet{nk12}, we simply
assumed a constant value of $\theta_s=5''$ for both sources. 
Through observing more than one transition from the same molecule, we can deduce its
$T_{\rm ex}$ and $N$ using a straight-line fit to ${N_u}/{g_u}$ versus ${E_u}/{kT_{\rm ex}}$.

\begin{figure*}
\epsfig{file=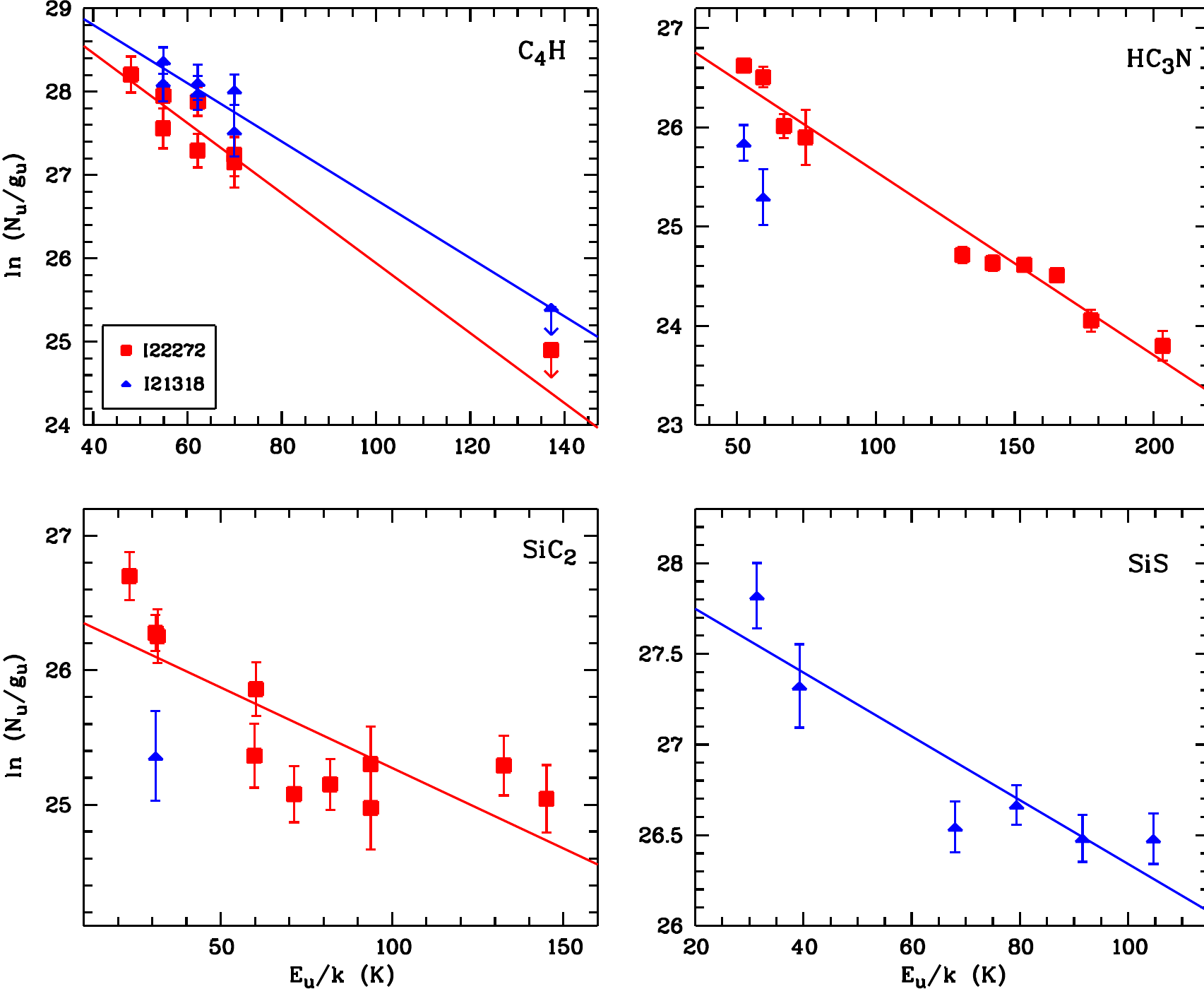,
height=12cm}
\caption{Rotation diagrams for the detected species.  }
\label{dia}
\end{figure*}

The rotation diagrams for the molecules detected in the two CSEs
are plotted in Figure~\ref{dia}. There is no significant difference
between the obtained $T_{ex}$ in the two sources. 
The resultant $T_{ex}$ and $N$ (or the upper limits) values are given in Table~\ref{abu}.
The errors were estimated from the uncertainties of the measurements and fittings.
We find $T_{ex}$(SiC$_2$)$>T_{ex}$(SiS)$\simeq T_{ex}$(HC$_3$N)$>T_{ex}$(C$_4$H)$>T_{ex}$(CS).
For both sources, $T_{ex}$(CS) is much lower than the other $T_{ex}$ values,
probably attributing to the large dipole moment of CS. Some molecules do not
have enough transitions to construct the rotation diagrams. We calculated their
column densities by assuming the $T_{ex}$ values as given in Table~\ref{abu}.
It should be noted that for optically thick lines the obtained $N$ values represent
only lower limits.

The molecular fractional abundance relative to H$_2$, f$_{\rm X}$, can be calculated
by dividing the molecular column densities by that of H$_2$. However, to the author's
knowledge, there is no published report on the detection of H$_2$ lines in the two sources. 
\citet{ds03} have failed to detect H$_2$ emission in IRAS\,$22272+5631$. We thus
simply assumed a typical value of f$_{^{13}\rm CO}=10^{-4}$ for the abundance calculations.
Although this assumption is somewhat arbitrary, the introduced uncertainties can be 
reduced when comparing the relative molecular abundances.
Table~\ref{abu} shows the obtained f$_{\rm X}$ values and the abundance upper limits for 
non-detected molecules.

\subsection{Isotopic ratios}

The isotopic ratios in CSEs can be altered by the products of nucleosynthesis 
in the stellar interior and drudge-up processes during the AGB phase.
According to AGB nucleosynthesis theory, $^{13}$C in CSEs can be greatly enhanced
\citep[see, e.g.,][]{mil09}, while the isotopic composition of sulfur is hardly
affected. We detected the isotopologues $^{13}$CO, C$^{34}$S, and H$^{13}$CN, allowing us
to estimate the isotopic ratios of carbon and sulfur, and to investigate the
nucleosynthetic histories and mixing processes of the two sources.
However, as the main lines, CO, CS, and HCN, are likely to be optically
thick, we can only determine the lower limits of the $^{12}$C/$^{13}$C and
$^{32}$S/$^{34}$S ratios, as given in Table~\ref{isot}. The much lower lower-limits of 
 $^{12}$C/$^{13}$C with respect to the solar value are unlikely to be fully attributed
to the effect of optical depth, and thus indicate the enhancement of $^{13}$C due to
nonstandard mixing processes. No difference between the $^{32}$S/$^{34}$S ratios
in the two IRAS sources, IRC+10216, and the Sun was found, consistent with
the prediction of AGB nucleosynthesis theory. 

The standard AGB stellar model predicts that
$^{14}$N in CSEs can be enhanced through the reactions $^{13}$C($p,\gamma$)$^{14}$N and 
$^{17}$O($p,\alpha$)$^{14}$N.  This is supported by the observation that the 
$^{14}$N/$^{15}$N ratio in IRC+10216
is much  higher than the solar value. Although the HC$^{15}$N lines in our spectra
are below the detection limit, we can roughly estimate the lower limit of the 
$^{14}$N/$^{15}$N ratios based on the H$^{13}$C$^{14}$N/H$^{12}$C$^{15}$N lower limit and 
the assumption that the $^{12}$C/$^{13}$C ratios are the same as that in IRC+10216.
Our results show that the $^{14}$N/$^{15}$N ratios in the two sources appear to be
higher than the solar value, supporting the enhancement hypothesis in AGB stars.
The $^{16}$O/$^{17}$O ratio can also provide insight into the AGB 
nucleosynthesis history since
 $^{17}$O can be enhanced by the CNO cycle reactions and $^{16}$O can be 
consumed through hot bottom burning in AGB stars. 
Unfortunately, no C$^{17}$O line lies within the observed frequency ranges.
In summary, we did not find any anomalous isotopic ratio in the 21 and 30\,$\mu$m sources.

\section{Discussion}

\subsection{Cyanopolyyne chemistry }

In the spectra of IRAS\,22272+5435, CN manifests itself as six individual peaks at
the 1.3\,mm window (Figure~\ref{f2}), which can be divided into three fine-structure 
groups $J=3/2$--3/2, $J=3/2$--1/2, and $J=5/2$--3/2 around 226.3, 226.6 and 226.8\,GHz.
We find that the intensity ratio between the 226.8\,GHz and the 226.6\,GHz groups
is smaller than 1, significantly different from the  intrinsic strength ratio of 1.8.
The anomalous CN intensity ratios  have been reported in C-rich and O-rich AGB stars 
\citep{bf97b}. If one attributed this to the effect of optical thickness, the required
line opacity would be unreasonably large. Therefore, following \citet{bf97b}, we
 interpret the observed anomalies as the pumping by optical and/or near-infrared radiation.

HCN and its geometrical isomer HNC are important reactants to produce cyanopolyynes.
Although both can be formed through a similar pathway, HCN has a lower energy and 
thus is more table than HNC. At certain physical conditions, they can convert to each
other.  Therefore, the HNC/HCN abundance ratio can reflect the physical conditions
that induce the production and isomerization of the two molecules.
\citet{ja84} found the HNC/HCN ratio in IRC+10216 to be 0.004, significantly lower 
than those found in dark cloud cores ($>1$). This could be attributed to the very 
limited region in the CSE over which HNC is formed  \citep{nm87}.
Our calculations show that the HNC/HCN ratios in IRAS\,21318+5631 and 22272+5435 
are lower than 0.14 and 0.22, respectively. 
If assuming that the H$^{12}$CN/H$^{13}$CN  abundance
ratio is equal to the $^{12}$C/$^{13}$C ratio in IRC+10216, we obtain the
HNC/HCN ratios of 0.0085 and 0.0081 in IRAS\,21318+5631 and 22272+5435, respectively.
These values are similar to that of IRC+10216.  Higher HNC/HCN ratios
were observed in PNs \citep[with an average of 0.5;][]{bf97}, indicating an ongoing chemistry
during the AGB-PPN-PN transition.

Under the same assumption as above, we obtain the CN/HCN abundance ratios of
0.34 and 0.22 in  IRAS\,21318+5631 and 22272+5435, respectively. As 
CN can be significantly produced from HCN through photodissociation in the
PPN-PN phases,
the CN/HCN abundance ratio can be taken as an  indicator of the degree of
CSE evolution \citep{bf97b}.  The CN/HCN abundance ratio in IRC+10216 is about 0.12
\citep{db95}, significantly lower than the average value of 9 in PNs \citep{bf97}.
This reflects the photodissociation from HCN into CN caused by the increasing temperature 
of the central sources. The CN/HCN ratios of the two IRAS sources are close to
that of IRC+10216, suggesting that the radiation fields of their central stars are not hard enough to
substantially affect cyanopolyyne chemistry.

HC$_3$N is less abundant than CN and HCN in both sources. 
Because HC$_3$N in CSEs can be produced by CN and/or HCN via neutral-neutral reactions,
the HC$_3$N/CN and HC$_3$N/HCN abundance ratios can serve as a tracer of circumstellar 
evolution. Our calculations show that the HC$_3$N/CN and HC$_3$N/HCN abundance ratios in
IRAS\,22272+5435 are higher than those in IRAS\,21318+5631, indicating to an enhancement
of longer cyanopolyyne molecules during the post-AGB evolution.

\subsection{Comparison with other CSEs}

The molecules formed in CSEs depend on not only the parent molecules ejected from inner zones, but also the varying environments driven by various processes such as mass loss, dust condensation, shocks, UV radiation, and so on. Theoretically, a certain process can preferentially enhance or destruct certain molecules over others, and thus certain molecules can potentially server as tools to investigate the circumstellar environments. However, all the processes are coupled together, making it extremely challenging to identify the key one influencing the abundance of a certain molecule from observations. For that, large number statistics of various molecules in CSEs is critically required. 
As part of a long-term project of studying circumstellar chemistry,
our research group has carried out systematic millimeter-wavelength spectral line surveys of a CSE sample,
including AGB stars, PPNs, and PNs (cf. the reference in Section 1 ). In order to ascertain whether 
the spectra of gas-phase molecules
can be used to distinguish between the 21 and 30\,$\mu$m sources and normal C-rich CSEs, we 
compare the line intensities of the two IRAS sources
and other sources obtained in our previous observations. Since all the observations were performed 
utilizing  the same instrumental settings,  the systematic uncertainties are minimized when comparing 
these spectra. Our spectra cover a wide frequency range where lines arising from upper levels with
different energies are detected, enabling us to investigate the coupled effects of 
excitation conditions, optical depths, and molecular abundances on line intensities.

\begin{figure*}
\epsfig{file=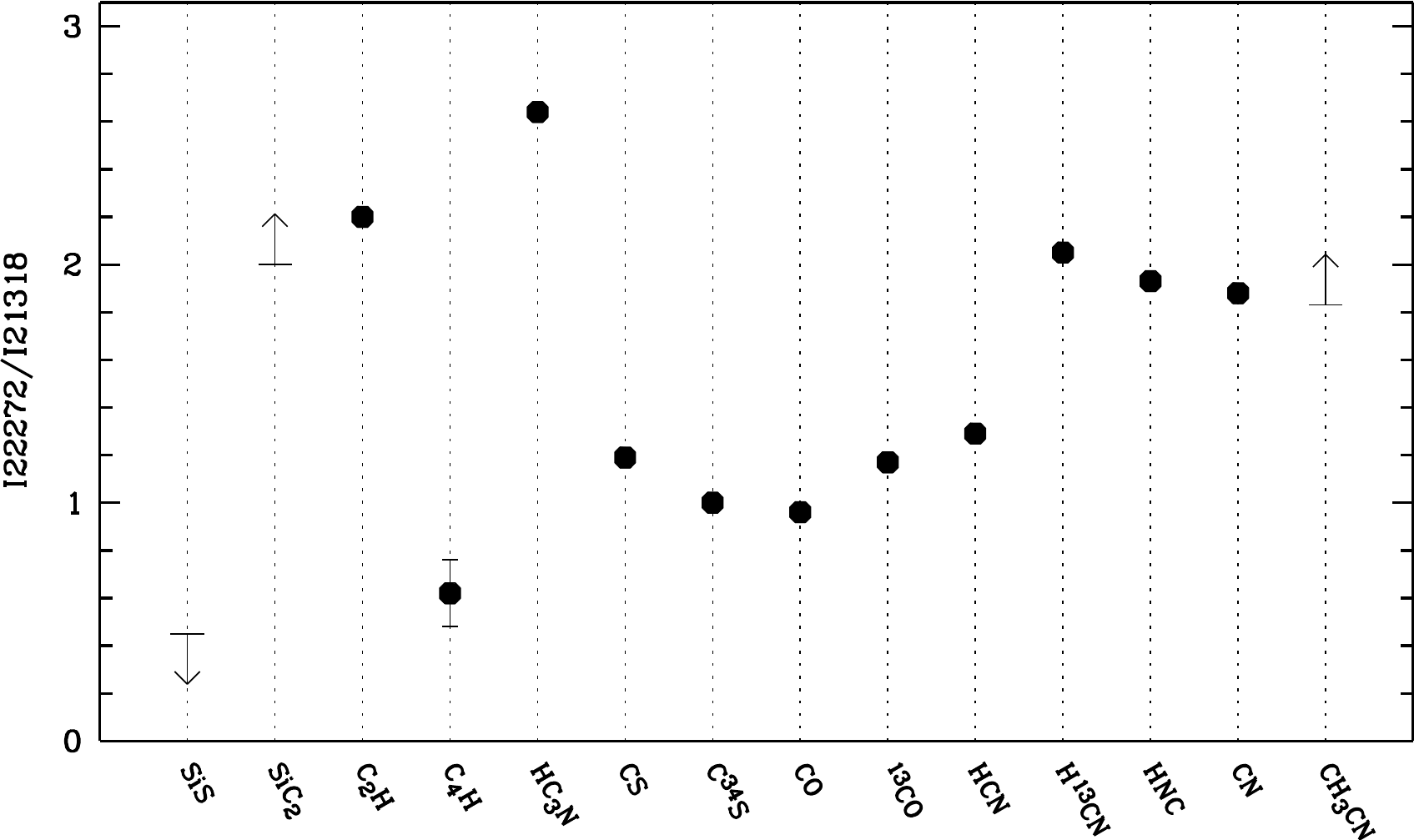,
height=9cm}
\caption{Integrated strength ratios of the lines detected in 
IRAS\,21318+5631 and IRAS\,22272+5435.  The error bars denote the standard deviations from the means
if more than one line are detected for a given species.
}
\label{compiras}
\end{figure*}

Figure~\ref{compiras} shows the intensity ratios of the lines detected in  IRAS\,21318+5631 and 22272+5435.
The latter clearly exhibits stronger SiC$_2$, C$_2$H, HC$_3$N, H$^{13}$CN, HNC, CN, and CH$_3$CN lines, and weaker 
SiS and C$_4$H lines. The generally stronger emission from carbon-bearing molecules indicates a
more C-rich environment in IRAS\,22272+5435. The different H$^{13}$CN/HCN ratios in the two sources can be in part attributed to the effects of optical depth.

\begin{figure*}
\epsfig{file=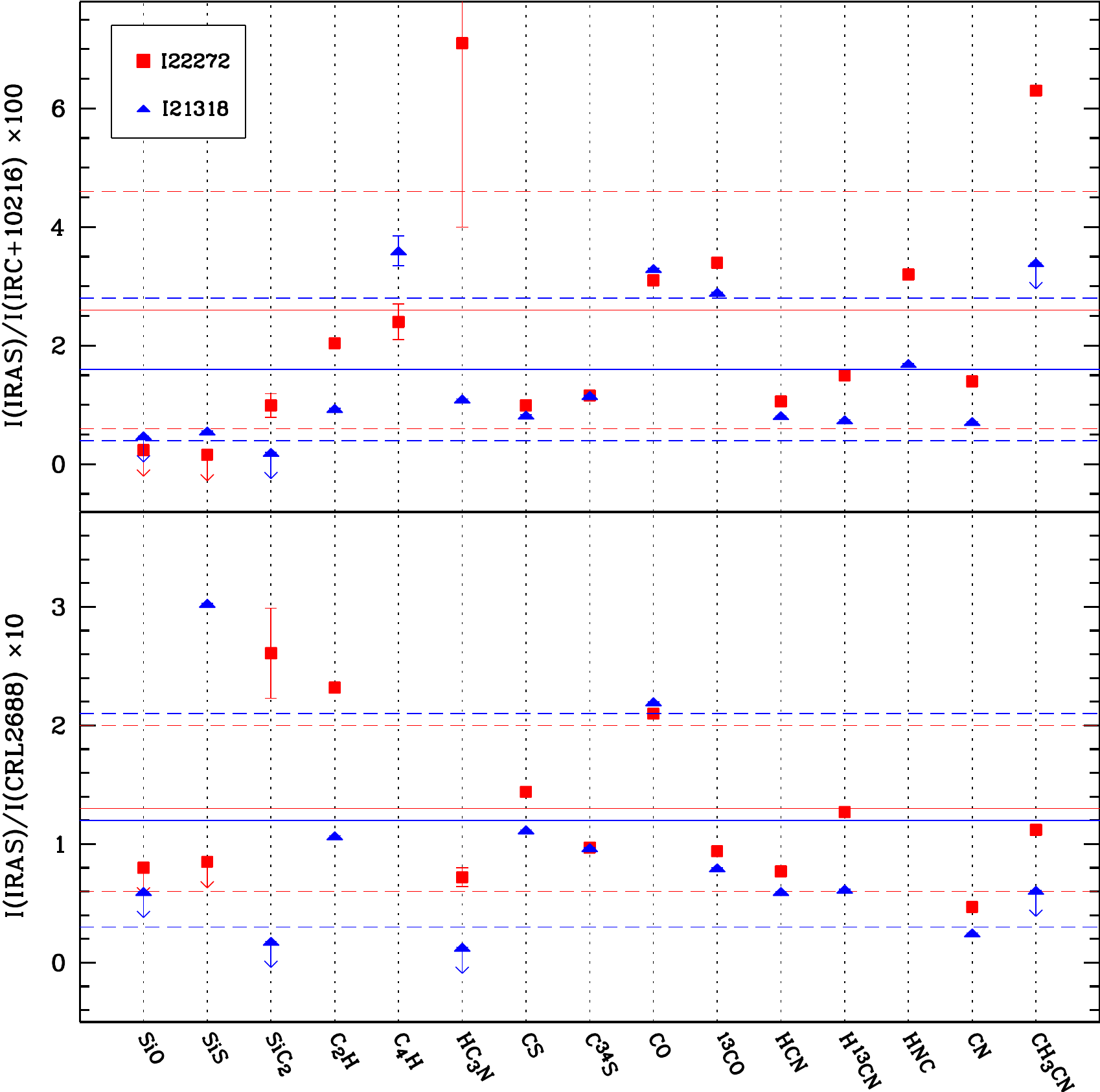,
height=15cm}
\caption{Integrated strength ratios of the lines detected in the
two IRAS sources and those detected in IRC+10216 and CRL\,2688. The error bars denote the standard deviations from the means
if more than one line are detected for a given species. The solid and dashed lines represent the mean values and standard 
deviations of the intensity ratios of all the molecular lines.
Note that the C$_4$H and HNC lines detected in the two IRAS sources lie out of the surveyed spectra range of CRL\,2688.
}\label{compothers}
\end{figure*}

In Figure~\ref{compothers} we compare the  intensity ratios of the lines detected in the
two IRAS sources and those in IRC+10216 and CRL\,2688. The millimeter spectra of the latter two CSEs are taken from \citet{he08} and \citet{zy13}. CRL\,2688 is a bright PPN that may be completing a transition between the `21\,$\mu$m phase' and the `normal phase' \citep{gt92}, and thus is the  immediate descendant of IRAS\,22272+5435.
We obtain the mean line-intensity ratios of IRAS\,21318+5621 and 22272+5435 over IRC+10216 
to be $(1.6\pm1.2)\times10^{-2}$ and  $(2.6\pm2.0)\times10^{-2}$, respectively, and those over CRL\,2688 to
be $(1.2\pm0.9)\times10^{-1}$ and  $(1.3\pm0.7)\times10^{-1}$, respectively.
The line intensities in the two IRAS sources are typically 10--100 times fainter than those in IRC+10216 and CRL\,2688, presumably due to their larger distances. Although the detected molecules in the two IRAS sources are much less than those in IRC+10216 and CRL\,2688, we cannot draw the conclusion that they are less abundant in chemistry.

An inspection of Figure~\ref{compothers} clearly shows that compared to IRAS\,21318+5621 and IRC+10216, IRAS\,22272+5435 and CRL\,2688
show a more similar pattern of the relative intensity ratios of molecular lines. This is compatible with the close evolutionary stages of  IRAS\,22272+5435 and CRL\,2688.
If taking the average intensity ratios as the reference, 
the two PPNs both exhibit stronger HC$_3$N and CH$_3$CN emission, and weaker SiS emission with respect to IRAS\,21318+562 and IRC+10216, suggesting that larger molecules are synthesized during the AGB-PPN transition. Nevertheless, there exist subtle differences between the
intensity ratio patterns of IRAS\,22272+5435 and CRL\,2688, with the former
showing enhancements of SiC$_2$ and C$_2$H. This might be a consequence of the extremely C-rich environment of this strong  21\,$\mu$m source.

Compared to IRAS\,22272+5435, the intensity ratio pattern of IRAS\,21318+5631 is more similar to that of IRC+10216, which can be told from the standard deviations of the mean intensity ratios
as shown by the dashed lines in Figure~\ref{compothers}.
Nevertheless, it exhibits relatively weaker emission from Si-bearing compounds, suggesting that this extreme carbon star has suffered from efficient dust condensation causing depletion of gas-phase Si-bearing molecules. The same properties have been found in CRL\,3068, a prototypical extreme carbon star in our sample \citep{zy09b}. 

%We do not find any abnormal line intensity ratio that is exclusively associated with the 30\,$\mu$m source.

\begin{figure*}
\epsfig{file=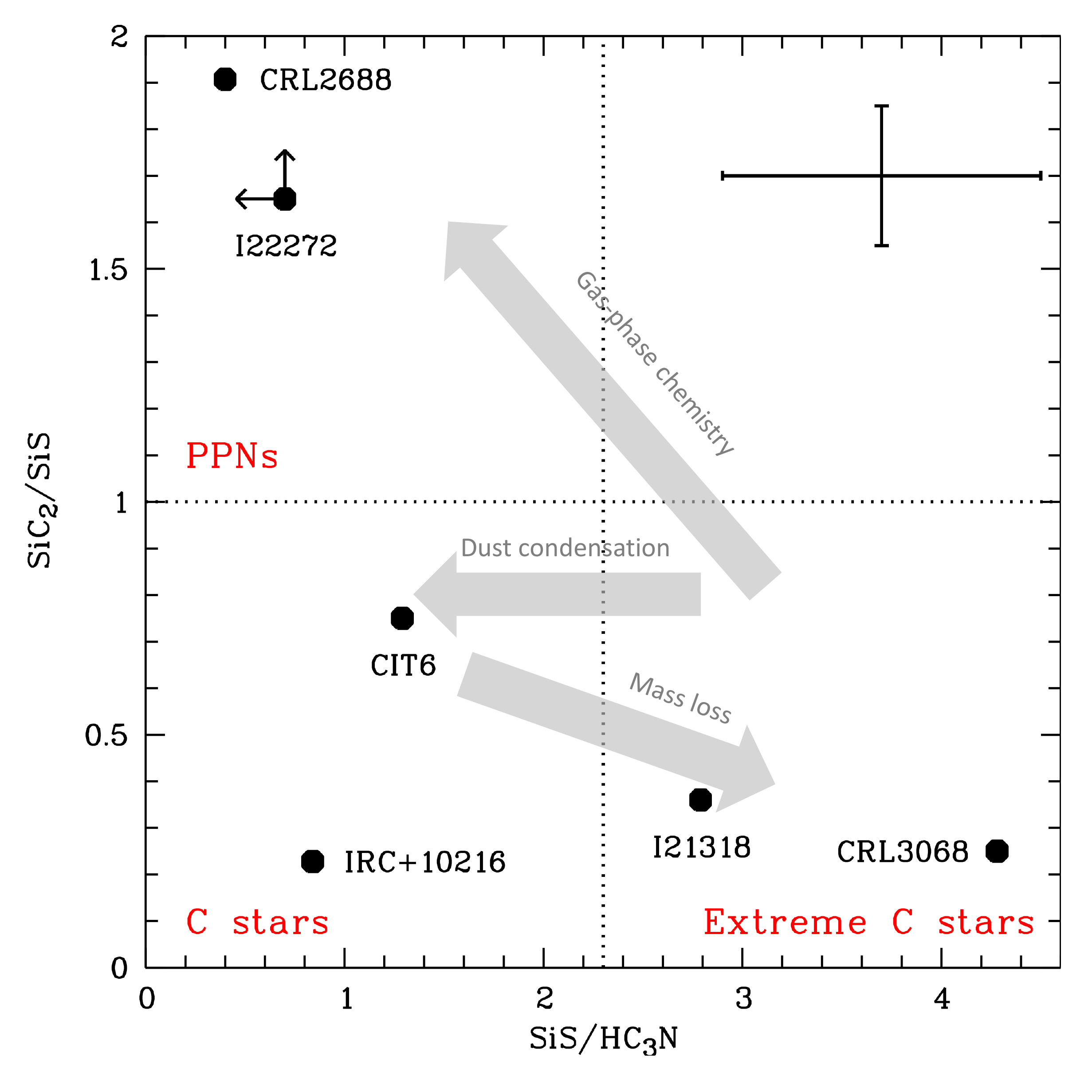, 
height=15cm} 
\caption{SiC$_2$/SiS and SiS/HC$_3$N abundance ratios determined for
a sample of CSEs. The typical error bar is shown in upper right corner.
The C-rich AGB stars, extreme carbon stars, and PPNs in 
this diagram can be separated by the dotted lines. The grey big 
arrows denote the probable evolutionary directions driven by the marked processes. 
}\label{abucom}
\end{figure*}

In order to obtain a useful indicator of the evolutionary stages of CSEs,
we examine the abundance ratios of various molecules detected in our sample. We find that
the CSEs in different  evolutionary stages can be well separated with the SiC$_2$/SiS and 
SiS/HC$_3$N abundance ratios, as shown in Figure~\ref{abucom}. During the evolution from
AGB stars to extreme carbon stars,  circumstellar SiS and SiC$_2$ have a photospheric origin, and  can be 
continually enhanced through stellar winds. When the extreme carbon stars reach the PPN
phase, the envelopes greatly expand and are detached from the central stars. Increasing
photochemical reactions cause the enhancement of SiC$_2$ and HC$_3$N. At the same time,
Si-bearing molecules are deposited on to dust grains because of their refractory nature. These processes can
take place in a timescale of $10^3$--$10^4$ years.
Such a picture of chemical evolution is roughly sketched in Figure~\ref{abucom}.

\subsection{Implications on the carriers of the 21\,$\mu$m and the 30$\mu$m
features}

All the molecules detected in  IRAS\,22272+5435 are C-bearing compounds,
and this 21\,$\mu$m source does not show SiS and SiO lines that are
usually observed in C-rich CSEs. \citet{hk91} detected
strong molecular bands of pure carbon molecules C$_2$ and C$_3$ in this source, 
 which might be the major precursor of carbon dust.
These observations demonstrate a remarkably C-rich environment of
IRAS\,22272+5435. The extreme C-rich nature can also be confirmed
by the high H$^{13}$CN/$^{13}$CO abundance ratio.
The HCN/CO line intensity has been found to be a good indicator of  the C/O \citep{ol93} 
in that HCN molecules are formed by the remaining carbon atoms that are not locked in CO.
 The abundance ratio of the less optically thick isotopologues,
H$^{13}$CN/$^{13}$CO, in IRAS\,22272+5435 is about two times larger than
that in IRAS\,21318+5631, indicating a higher C/O ratio.
It is therefore a natural speculation that the carrier of the 21\,$\mu$m feature 
is very likely to be carbon-dominant materials, such as carbon nanotubes and 
fullerene derivatives.

 In order to understand the environments favoring the formation and survival of the carrier of the 21\,$\mu$m feature, we should investigate the molecules not restricted to C-bearing ones.
\citet{sch07} found that the abundance of SiS in CSEs increases with increasing
C/O ratio, which can be explained in terms of photospheric LTE chemical models. In contrary to this scenario,  IRAS\,22272+5435 exhibits non-detection of SiS and a high C/O ratio. 
On the other hand, \citet{sch07} showed a tendency of decreasing SiS
abundance with increasing density of the wind, which reflects the freeze-out onto dust grains. 
The remarkably low expansion velocity \citep[$\sim10$\,km\,s$^{-1}$;][]{nk12} 
in IRAS\,22272+5435 indicates
a high density of the wind. As a result, we can conclude that SiS has been heavily adsorbed onto dust grains in this 21\,$\mu$m source.  Therefore, IRAS\,22272+5435 is characterized by high dust condensation efficiency,
 so that  surface reactions, such as hydrogenation of fullerene clusters, probably play an important role in the formation of the carrier of the 21\,$\mu$m feature.

 It is possible that the incorporation of Si into dust grains could produce solid state Si-bearing materials  emitting spectral signature at a position around 21\,$\mu$m, such as
silicon disulfide in the form of independent grains,  sulfurized SiC, and mantles on other
grains \citep{goe93}, silicon carbide with N or C impurities \citep{spc04}, and oxidized SiC \citep{pos04}.
However, it is hard to understand why these Si-bearing materials  are formed exclusively in
C-rich environments and can survive only in a very short time scale of PPNs.
Probably, the  Si-bearing materials and the carbon-dominant carrier of the 21\,$\mu$m feature have a core-mantle structure or stick together in an aggregate structure.

\citet{zha09c} and \citet{ll13} examined several Ti-, S-, Si-, and Fe-based
materials that have been proposed as the carrier candidates
of the 21\,$\mu$m feature, and found that
the available abundances of these elements are too low to be responsible for
the 21\,$\mu$m feature. The hydrides of fullerenes (fulleranes) have been proposed 
as the carrier of the 21\,$\mu$m feature \citep{web95,zhang20}. However, the emission peaks of fulleranes
shift from 19--20\,$\mu$m depending on the degree of hydrogenation.
This is inconsistent with the observations that the peak wavelengths
of this feature are nearly identical in different sources. Nevertheless, we cannot 
rule out the possibility that under special conditions, some of
fullerene derivatives or fulleranes,
probably those that are most resistant to UV radiation,
could be more favorably formed, and
a `magic' combination of these materials can contribute to the observed peak. 
\citet{zhang20} found that a tuned combination of fullerane isomers
is able to match the 21\,$\mu$m feature.
Fulleranes could be readily dehydrogenated by UV irradiation or shocks, compatible with the short survival time scale of the
21\,$\mu$m feature.
\citet{gt92} showed that
the 21\,$\mu$m feature sources also exhibit remarkably strong 3.4--3.5\,$\mu$m emission.
The features near 3.4--3.5\,$\mu$m can be in part attributed to
the C-H stretching modes of fulleranes \citep{zk13}, seemingly supporting the idea
that the 21\,$\mu$m feature is to some extent relevant to fulleranes.

The remarkably narrow line widths and strong 21\,$\mu$m feature in IRAS\,22272+5435 are rare among  CSEs.  
Is this behavior related to the 21\,$\mu$m  phenomenon? The slow expansion of the envelope 
can create a dense zone, where the dust formation  and solid surface reactions are efficient. 
Such an environment might favor the formation of the carrier of the 21\,$\mu$m feature. 
On the other hand, we reasonably conjecture that the carrier is so fragile that it could be readily destructed  by accelerated stellar winds.  
 \citet{so06} found that the abundance of SiO in C-rich CSEs is commonly higher than that predicted by equilibrium stellar atmosphere chemistry, and attributed this to a
shock-induced non-equilibrium chemistry. As a result, the non-detection of SiO in IRAS\,22272+5435 probably indicates the absence of shocks, which might be a necessary condition for the survival of the carrier of the 21\,$\mu$m feature. 

 Regarding to the carrier of the  30\,$\mu$m feature, we could hardly draw valid conclusions from our observational results. It is conceivable that the carrier of the 30\,$\mu$m feature 
should be negatively correlated with its gas-phase precursor in abundance.
IRC+10216 is a stronger 30\,$\mu$m source than IRAS\,21318+5631 and IRAS\,22272+5435 \citep{hon02}. Figure~\ref{compiras} shows that Si-containing molecules in the two IRAS sources are much more depleted than those in  IRC+10216, seemingly suggesting that  Si-containing molecules are unlikely to be the gas-phase precursor of the carrier of the 30\,$\mu$m feature. \cite{be94} and \citet{hon02} presented plausible evidences supporting MgS as the carrier of the 30\,$\mu$m feature. If this was
the case, the strength of the  30\,$\mu$m feature would be dominantly governed by the abundance of S because Mg is more abundant than S. According to the chemical equilibrium calculations of
 CSEs with a high C/O ratio \citep{agu20}, carbon is first condensed in the  most inner regions followed by SiC and then
 a variety of Si-containing minerals, while the condensation of MgS occurs in very outer regions.
The theoretical study of \citet{zh08} suggested that precipitation on SiC grains is the most promising mechanism of solid MgS formation, and if no SiC core is formed beforehand, the MgS condensation
is inefficient until gas-phase SiS molecules are significantly converted into solid SiC.
If  the 30\,$\mu$m feature arises from MgS, what is its possible gas-phase precursor? The abundance and direct condensation of gas-phase MgS are negligible in C-rich CSEs. 
\citet{goe85} suggested that MgS can be efficiently synthesized through the surface reaction of Mg and H$_2$S, although H$_2$S is relatively underabundant in C-stars. Through observational studies of CSE samples, \citet{sn12} and \citet{ma19}, respectively, found that there are anti-correlations between the strength of the 30\,$\mu$m feature and the abundances of SiS or CS,  and thus concluded that condensed MgS  is formed from gas-phase SiS or CS molecules.
Further radio observations of more gas-phase  molecules in larger samples could help to identify the gas-phase precursor of the carrier of the 30\,$\mu$m feature.

Although the MgS hypothesis has received relatively little debate, \citet{zha09d} argued that the available
MgS mass is insufficient to be responsible for the the 30\,$\mu$m  feature. \cite{lom12} reported that
the `mass problem' can be solved if MgS is present on the out layer of grains and has thermal contact
with amorphous carbon and SiC. Nevertheless, the candidate carriers proposed by far do not provide an explanation for the observational results that all 21\,$\mu$m emitters exhibit the 30\,$\mu$m  feature but there is no strict correlation between the strengths
of the two features. Laboratory spectra suggested that the amorphous hydrocarbon with mixed $sp^2$/$sp^3$ bonds can emit a prominent feature around 30\,$\mu$m \citep{gt01}. On the other
hand, \citet{zk13} and \citet{ot14} found that fullerene-containing PNs
and PPNs generally exhibit strong 30\,$\mu$m feature, suggesting that
photochemical processing of amorphous hydrocarbons might lead to the fullerene formation \citep{gar10,mic12}. 
 Experiments have revealed that fulleranes such as C$_{60}$H$_{36}$ can be rapidly synthesized through mixing
fullerenes and hydrogen atoms \citep{ig12}, while moderately hydrogenated fullerenes can
produce strong bands near 21\,$\mu$m \citep{web95,zhang20}. As such, we propose a highly speculative conjecture that the  21\,$\mu$m and  30\,$\mu$m features might be related to amorphous hydrocarbons. The  30\,$\mu$m feature origins from the
amorphous hydrocarbons  \citep{gt01}, or alternatively from the thermal contact of MgS and amorphous hydrocarbons \citep{lom12}. 
During the evolution of CSEs, once conditions are favorable, the amorphous hydrocarbons would partially form the fullerane isomers emitting the 21\,$\mu$m feature, which would quickly vanish due to UV- or shock-induced dehydrogenation.
This seems to offer a plausible explanation for the rarity of the 21\,$\mu$m phenomenon and
the observational correlation between the
21 and 30\,$\mu$m sources.

\section{Conclusions}

We present 2\,mm and 1.3\,mm observations towards an extreme carbon star harbouring the 30\,$\mu$m feature, IRAS\,21318+5631,  and a PPN harbouring the 21 and 30\,$\mu$m features, IRAS\,22272+5435. 
Our main goal is to investigate the relationship between the carriers of the 21 and 
30\,$\mu$m features and gas-phase chemistry, as well as to search for their gas-phase
building blocks.
Although the infrared spectra of the two IRAS sources show unusual emission features, 
their radio spectra are generally consistent with those of usual C-rich CSEs. Almost all the detected 
species are carbon-based.  We did not detect any unexpected gas-phase molecules. There is no evidence 
showing that their isotopic ratios differ from other C-rich CSEs.

We compare the spectra of the two IRAS sources with those of other CSEs previously 
obtained by us using the same instrument settings.  The results show that the 30\,$\mu$m sources
do not exhibit unusual behavior of molecular spectra, and thus we cannot
distinguish 30\,$\mu$m sources from normal CSEs based upon radio spectra only.  
However, the 21\,$\mu$m source exhibits relatively 
narrower line widths, stronger emission from long carbon-chain molecules, and non-detection of SiS and SiO.
We infer that the formation of the carrier of the 21\,$\mu$m feature requires an extreme C-rich and dusty
environment, and tentatively assign it as fragile carbon-dominant compounds formed on surfaces of dust
grains,  such as fulleranes; while the carrier of the 30\,$\mu$m feature might be related to  amorphous hydrocarbons. 
Further observations of a larger sample  of 21 and 30\,$\mu$m sources are clearly needed to
reach firm conclusions.

It is shown that the millimeter spectra can serve as a useful indicator of the evolutionary 
stages of CSEs. The different molecule abundances in different CSEs may reflect 
the gas and dust processes, the alterations of chemical drivers, and/or different properties of 
their progenitor stars. Wide-frequency-coverage observations of a large sample of CSEs with diverse 
properties allow us to assign a `circumstellar spectral classification' in the radio window. 
The current study based on very small number statistics is a small step towards achieving this ultimate
goal.

\acknowledgments

The author wishes to thank an anonymous referee for helpful comments toward improving this paper.
I also thank Kwok Sun and Nakashima Junichi for fruitful discussions.
The 12\,m telescope and the {\it SMT} are operated by the Arizona Radio Observatory ({\it ARO}), 
Steward Observatory, University of Arizona. I are grateful to the {\it ARO} staff for their 
helps during the observing run. This work was supported by National Science Foundation 
of China (NSFC, Grant No. 11973099).

\begin{figure*}
\epsfig{file=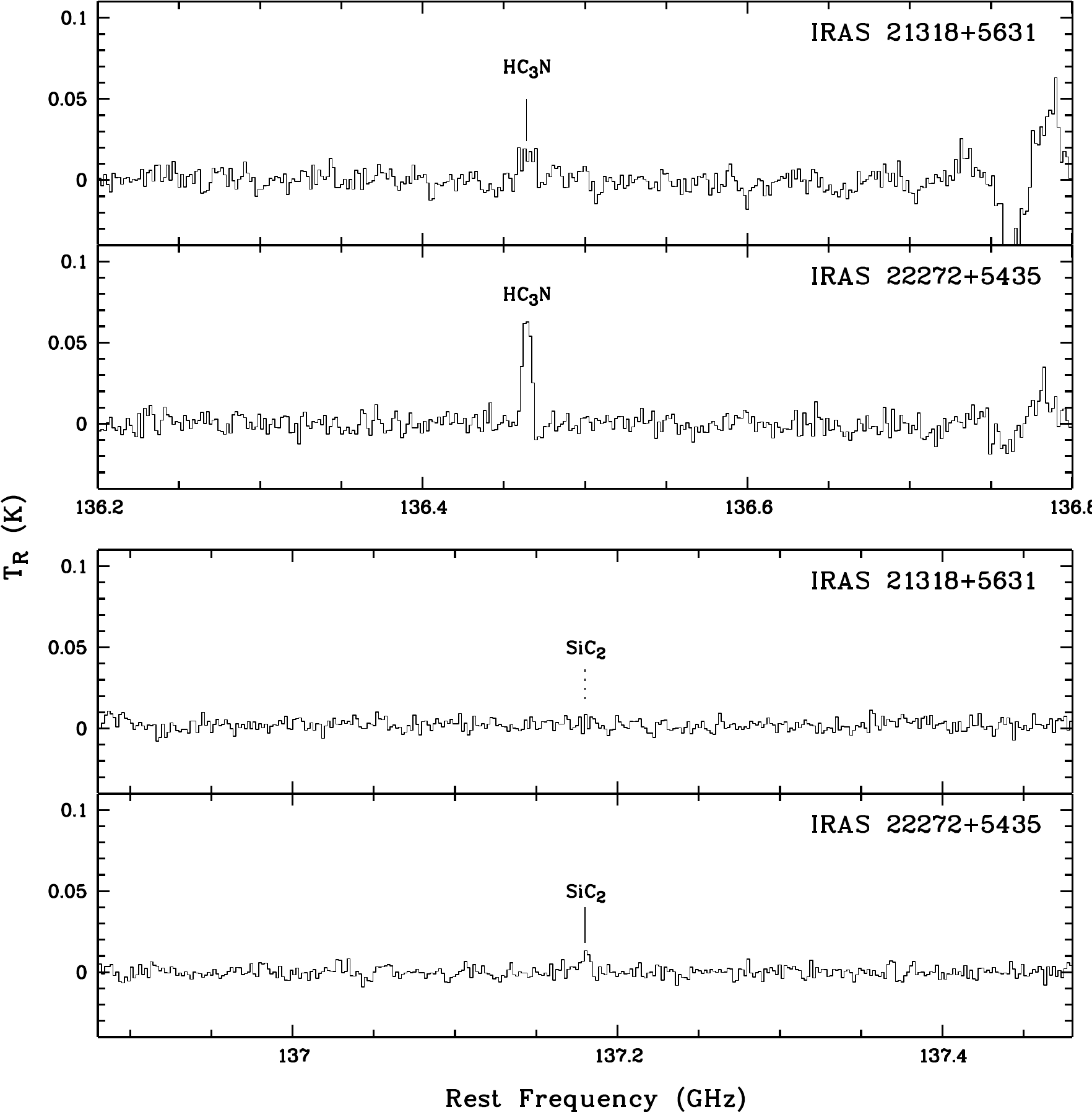,height=16cm}
\end{figure*}

\begin{figure*}
\epsfig{file=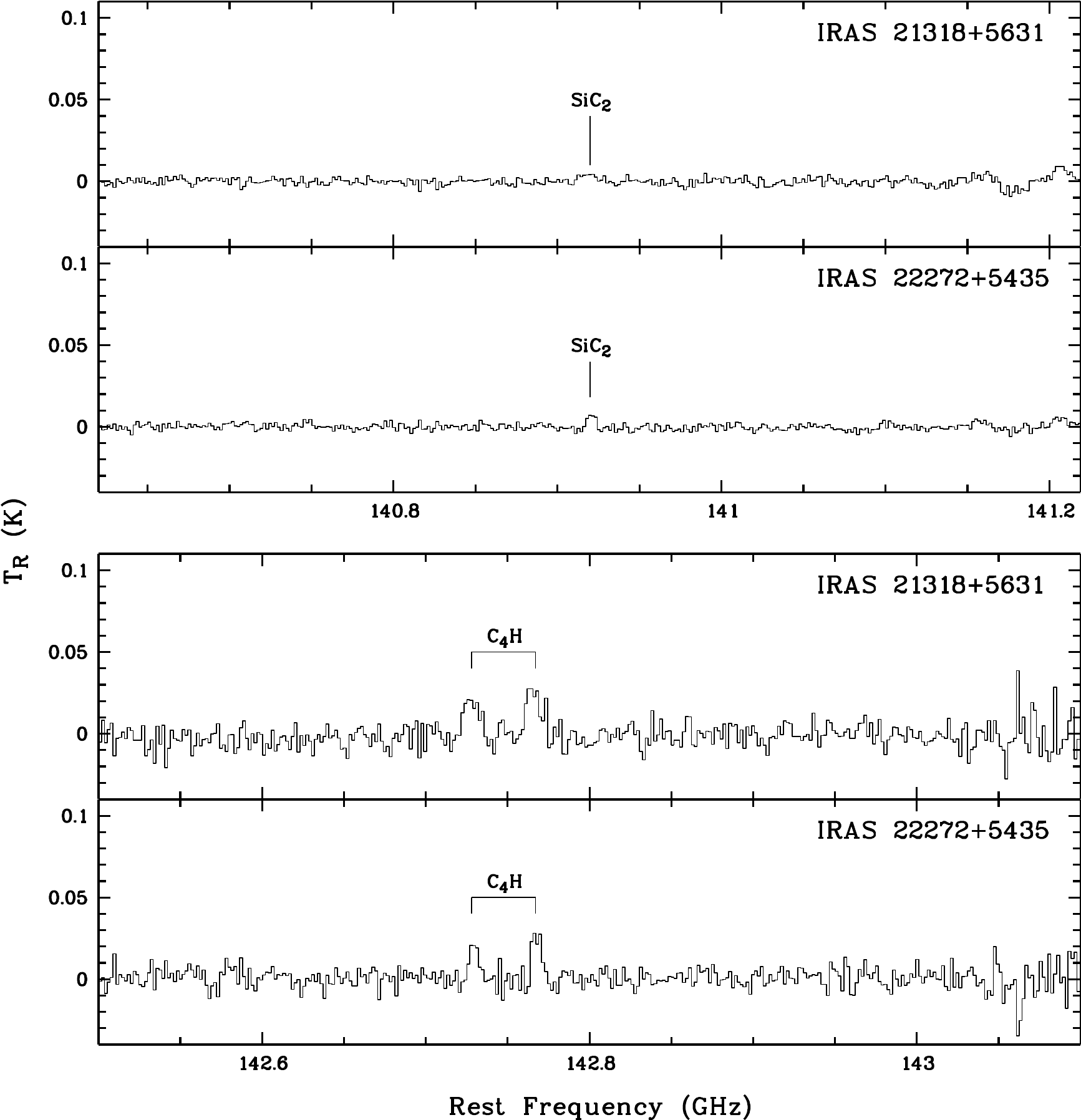,height=16cm}
\end{figure*}

\begin{figure*}
\epsfig{file=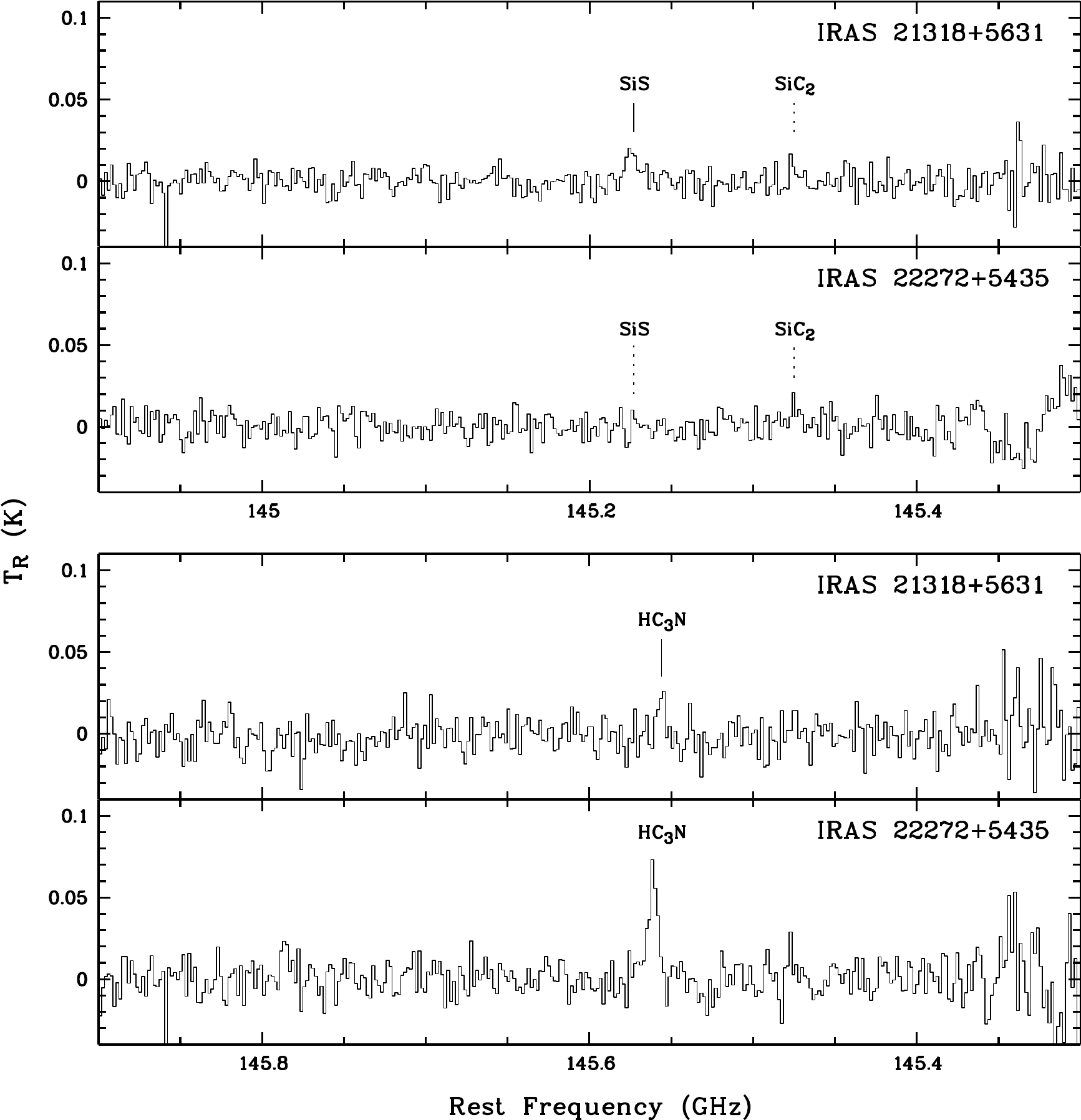,height=16cm}
\end{figure*}

\begin{figure*}
\epsfig{file=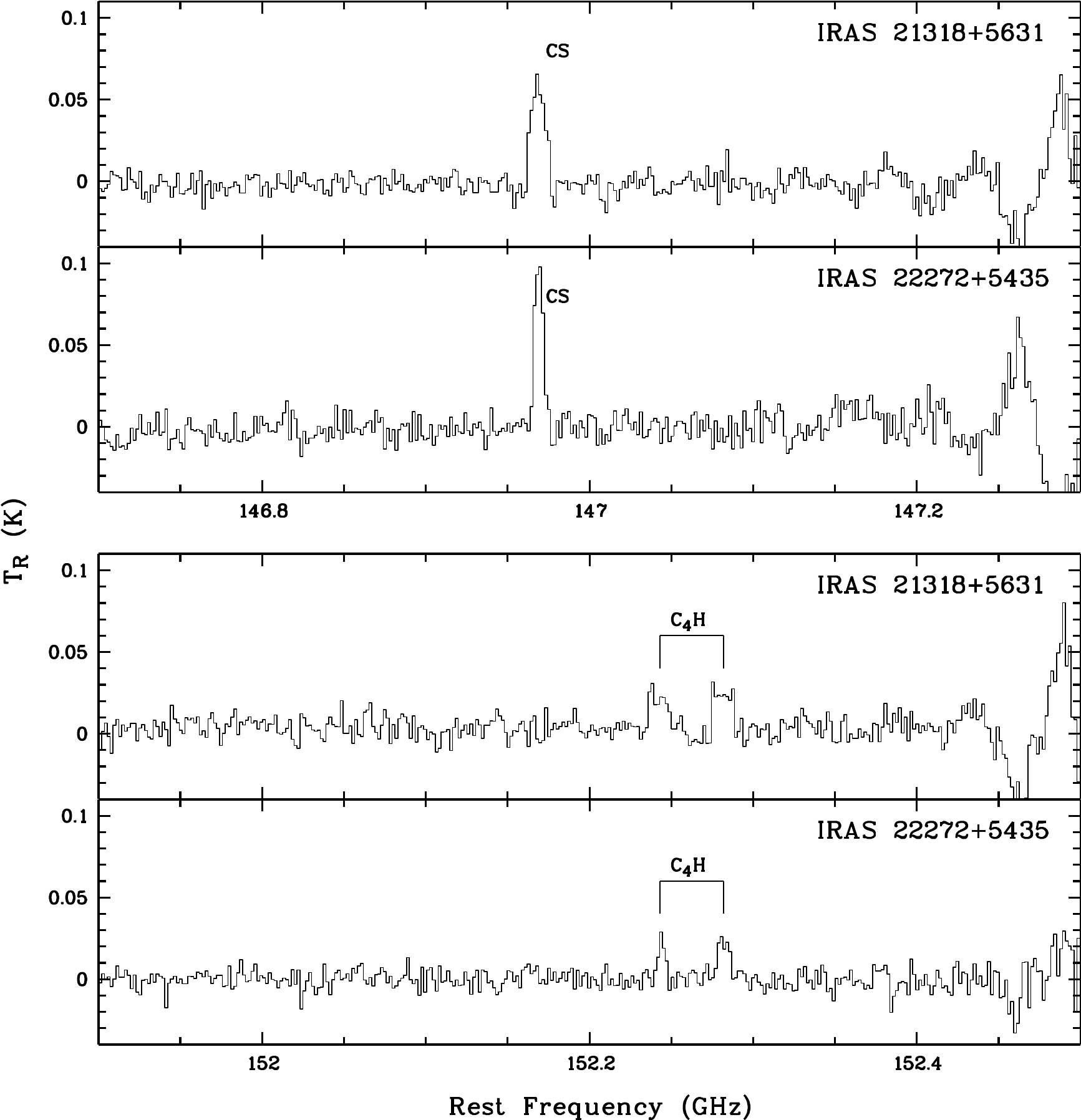,height=16cm}
\end{figure*}

\begin{figure*}
\epsfig{file=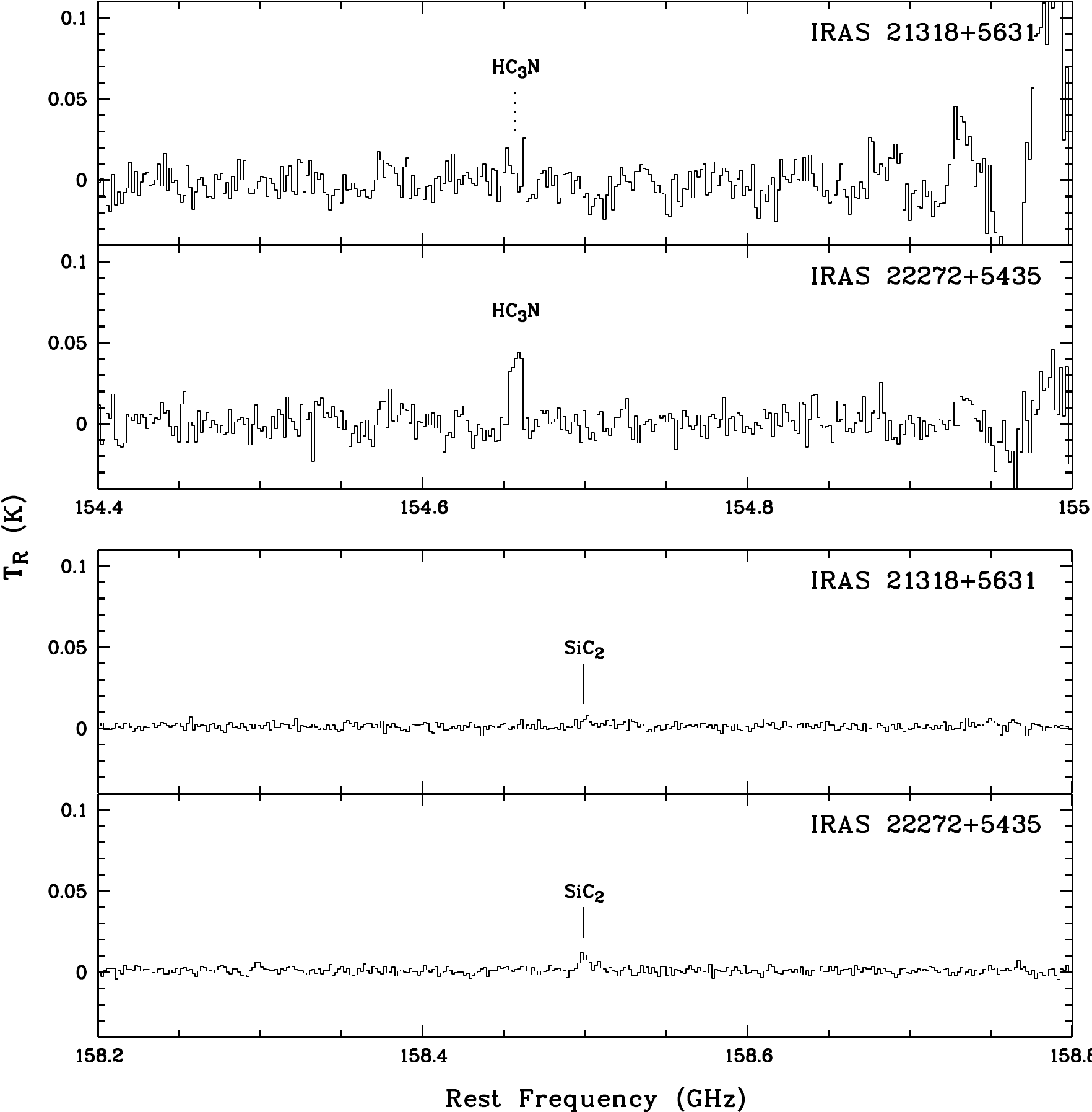,height=16cm}
\end{figure*}

\begin{figure*}
\epsfig{file=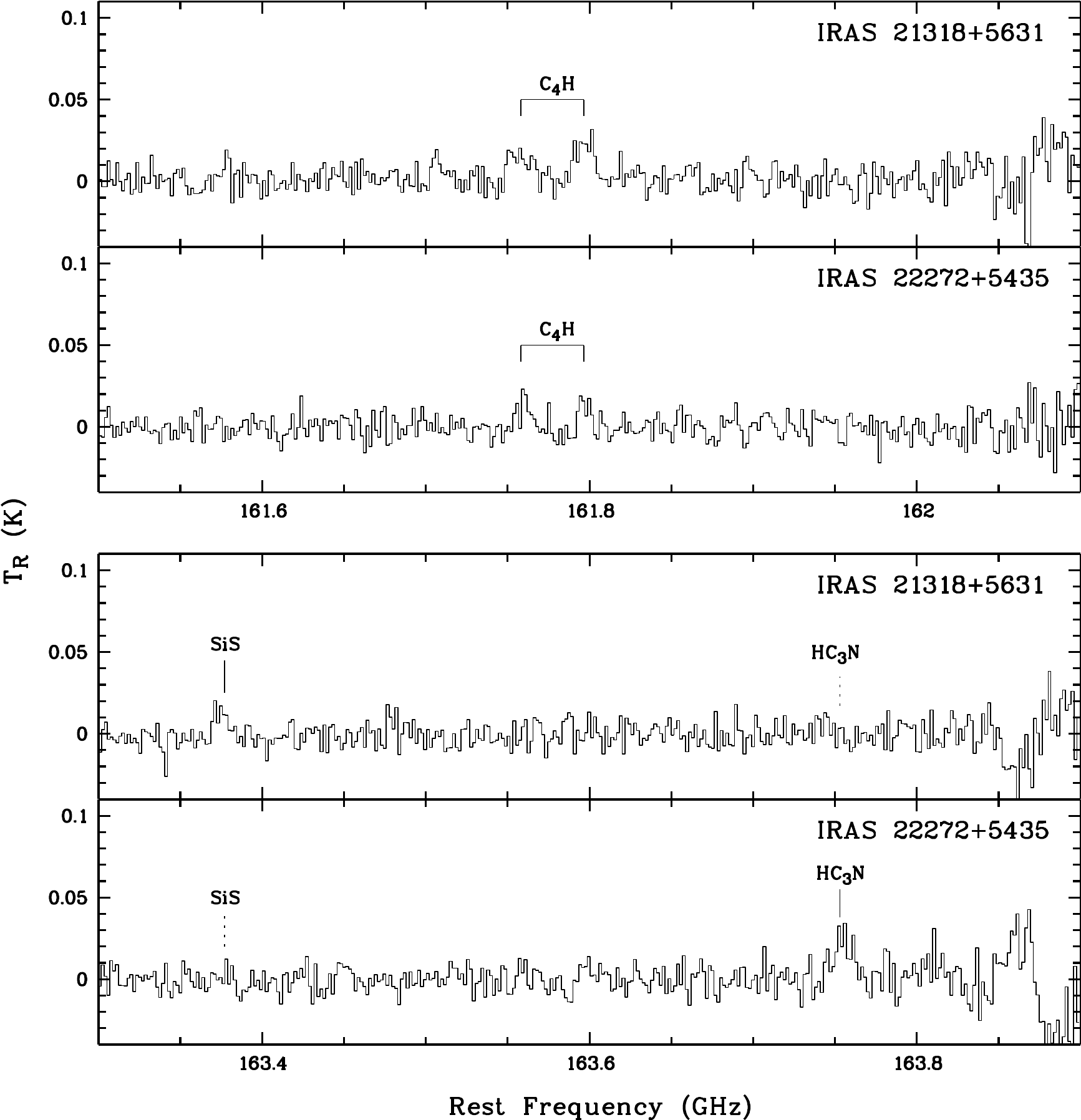,height=16cm}
\end{figure*}

\newpage
\clearpage

\begin{figure*}
\epsfig{file=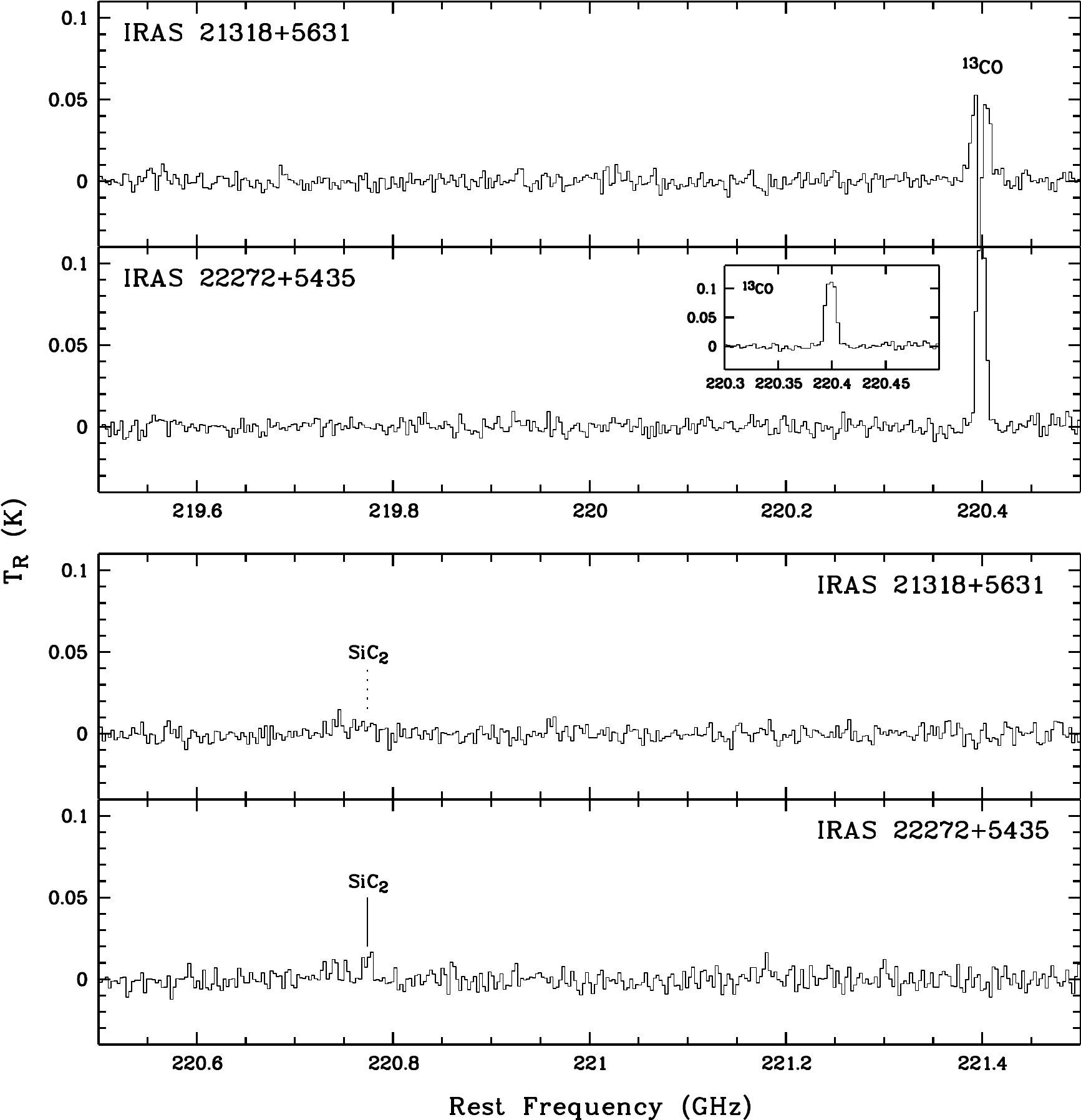,height=16cm}
\end{figure*}

\begin{figure*}
\epsfig{file=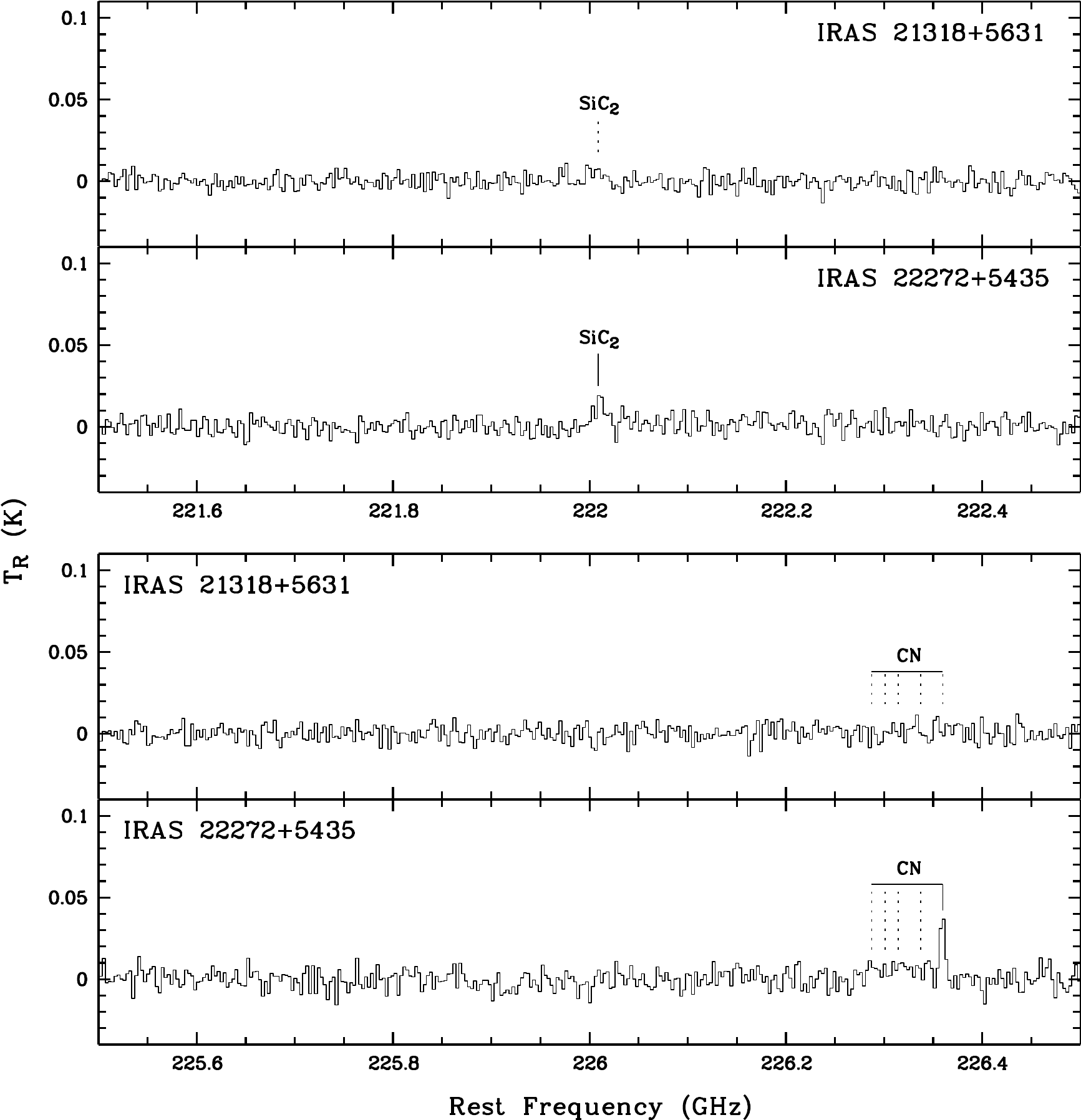,height=16cm}
\end{figure*}

\begin{figure*}
\epsfig{file=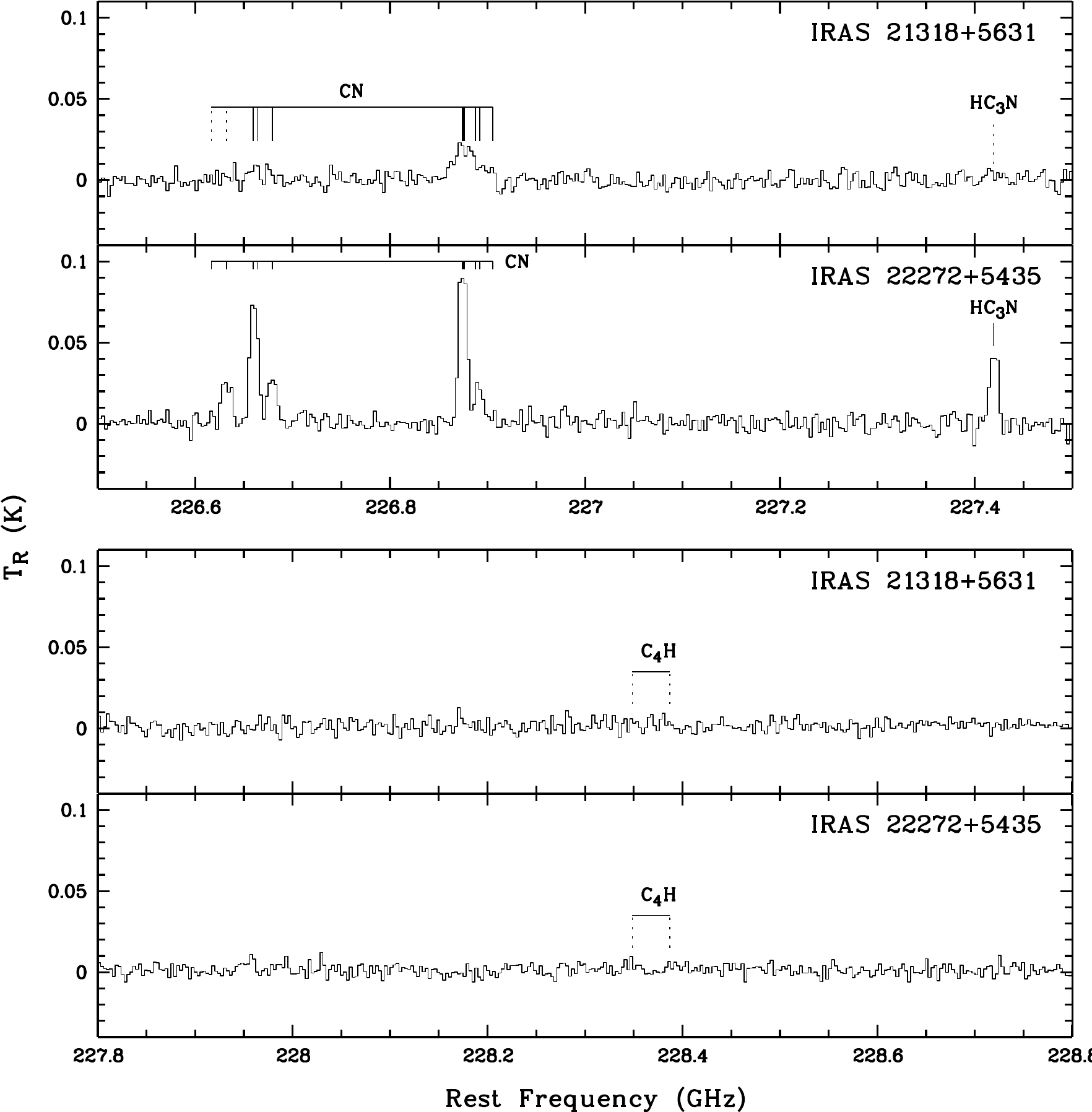,height=16cm}
\end{figure*}

\begin{figure*}
\epsfig{file=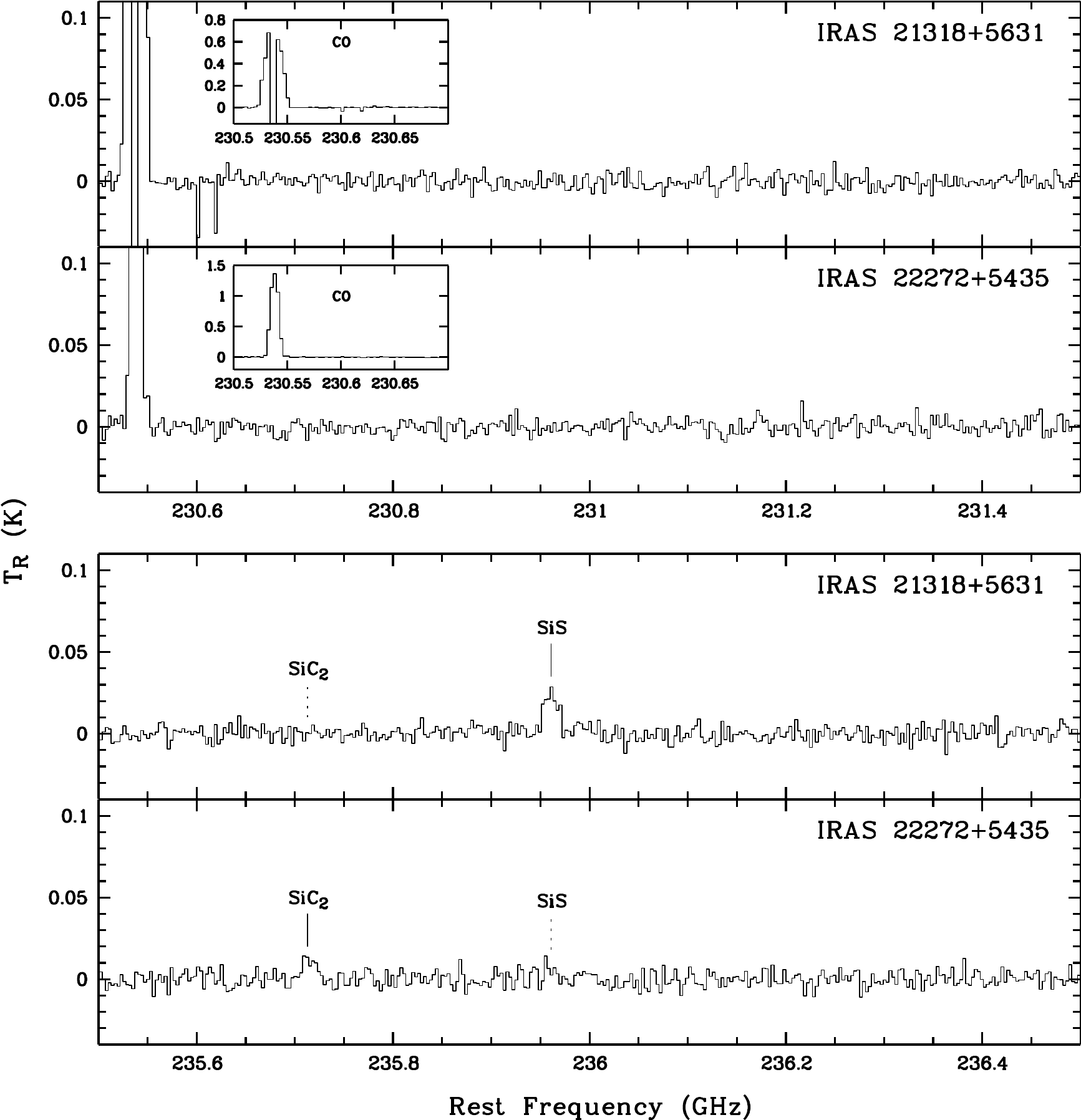,height=16cm}
\end{figure*}

\begin{figure*}
\epsfig{file=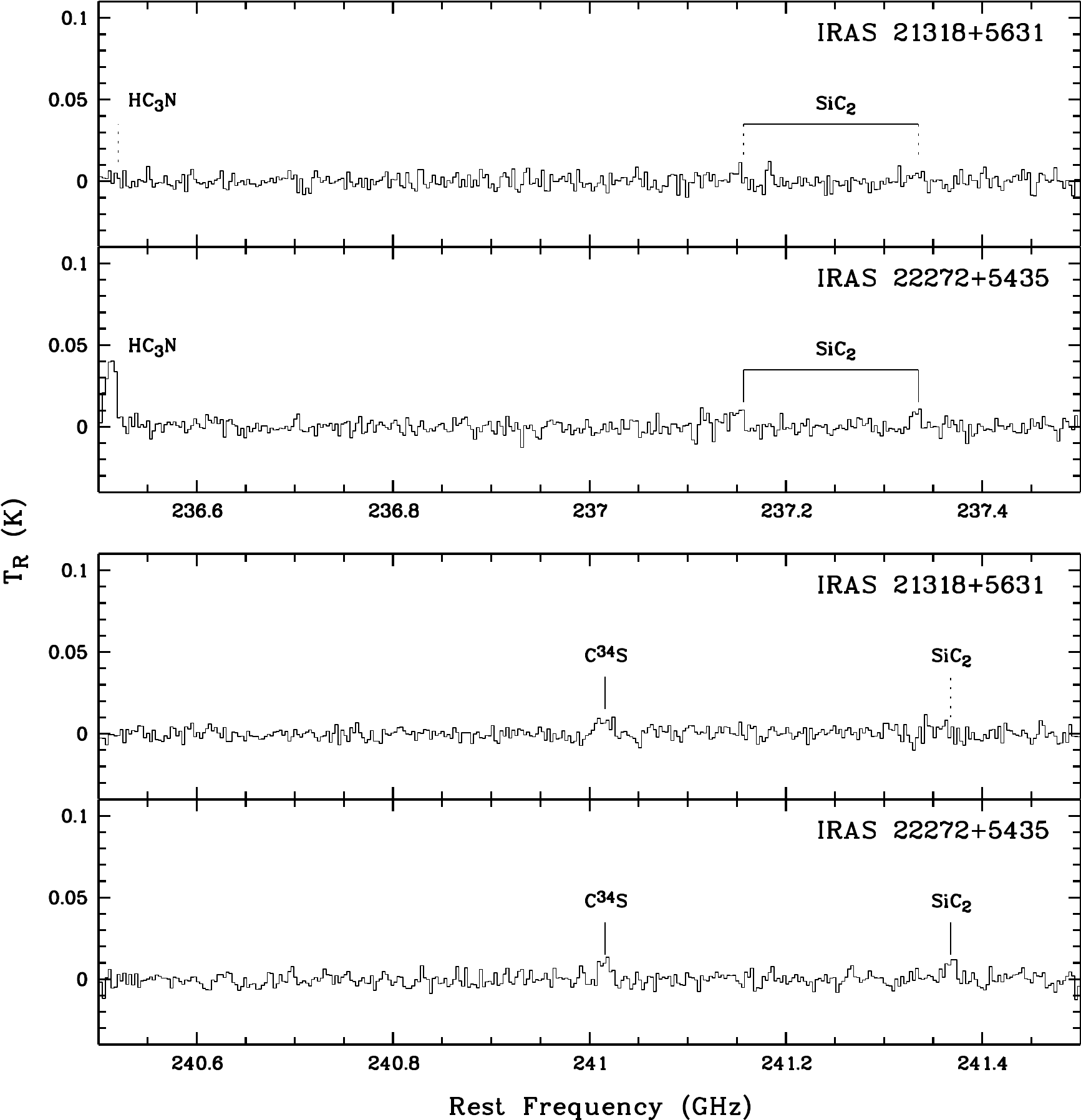,height=16cm}
\end{figure*}

\begin{figure*}
\epsfig{file=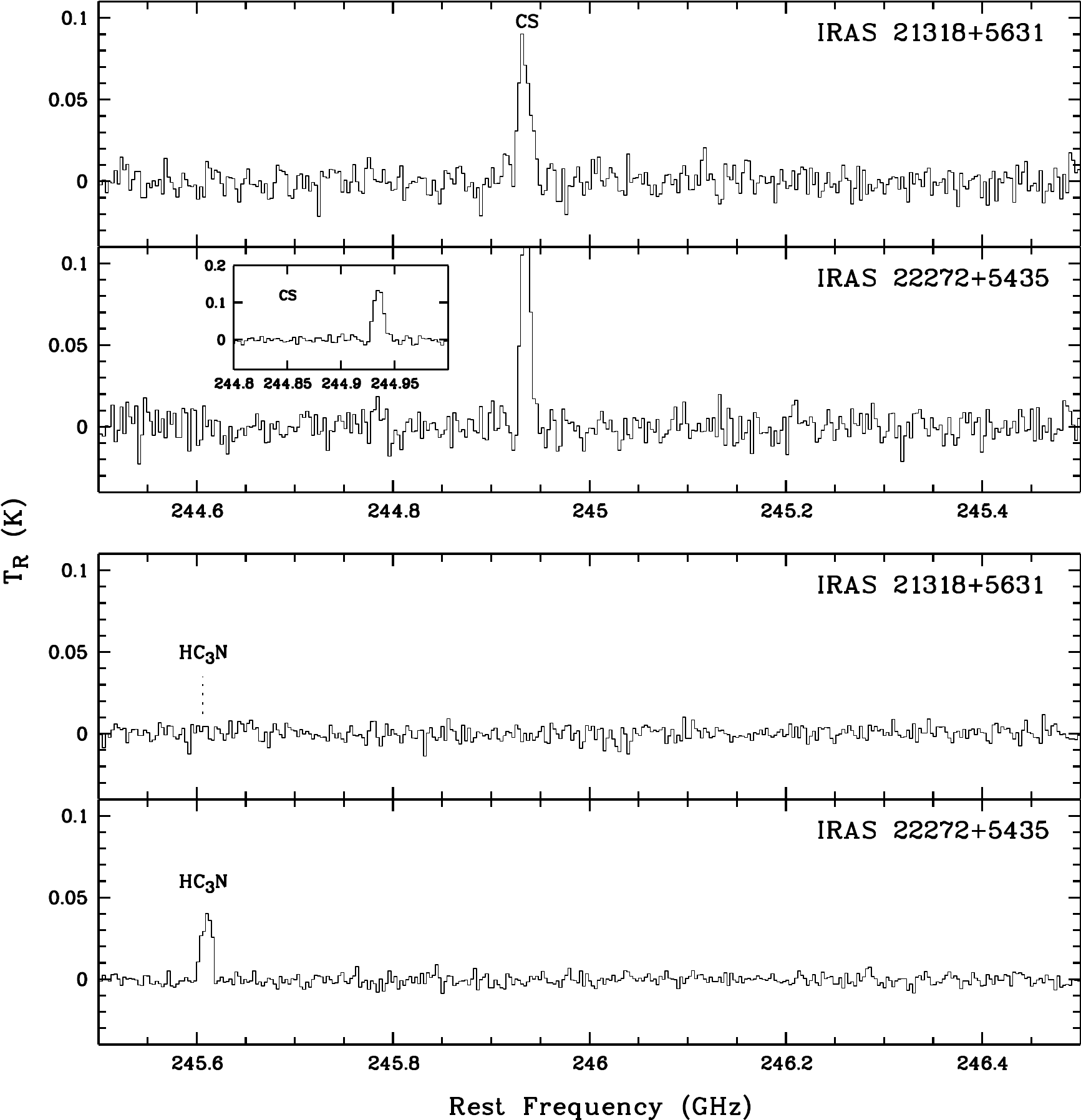,height=16cm}
\end{figure*}

\begin{figure*}
\epsfig{file=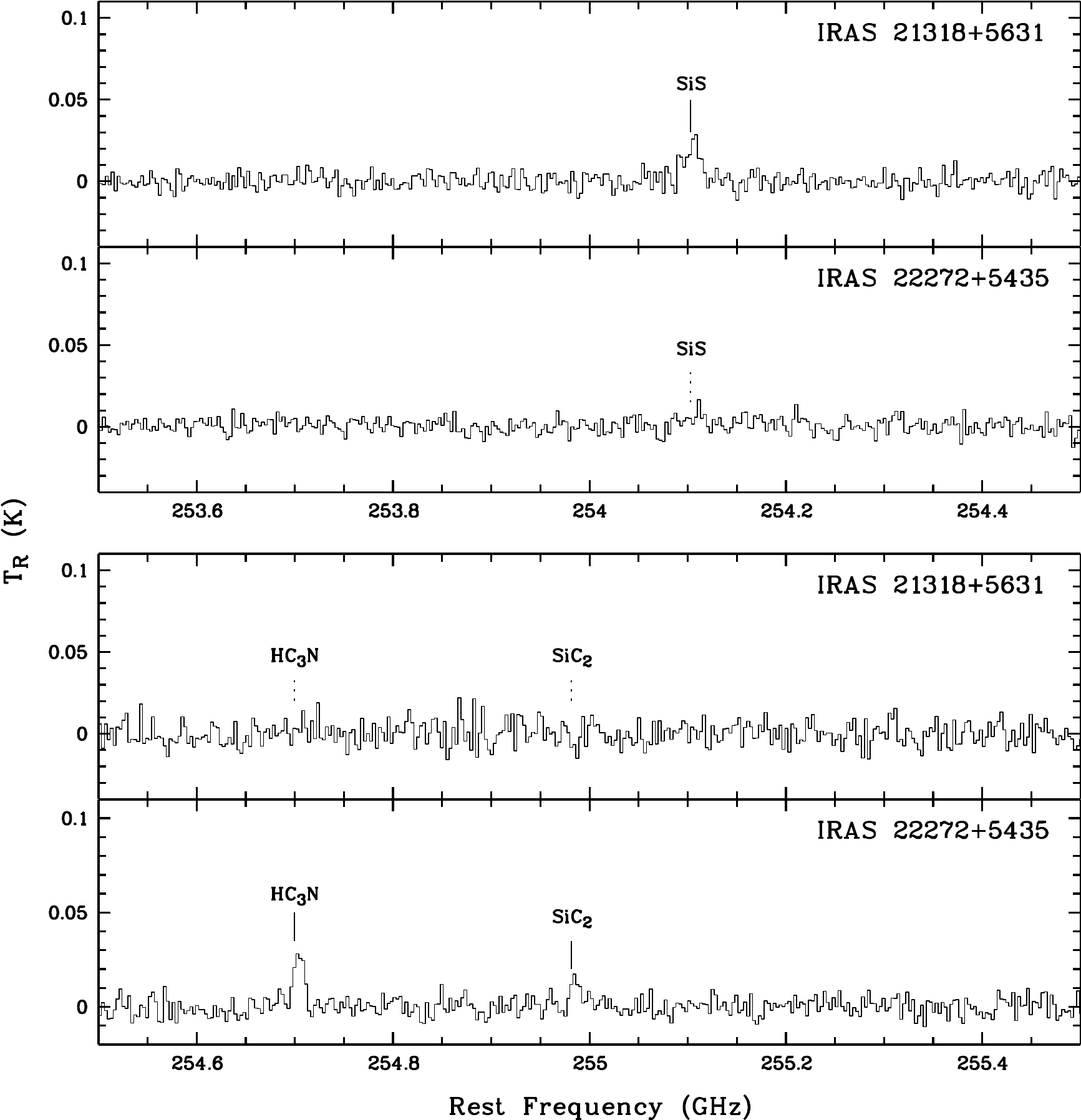,height=16cm}
\end{figure*}

\begin{figure*}
\epsfig{file=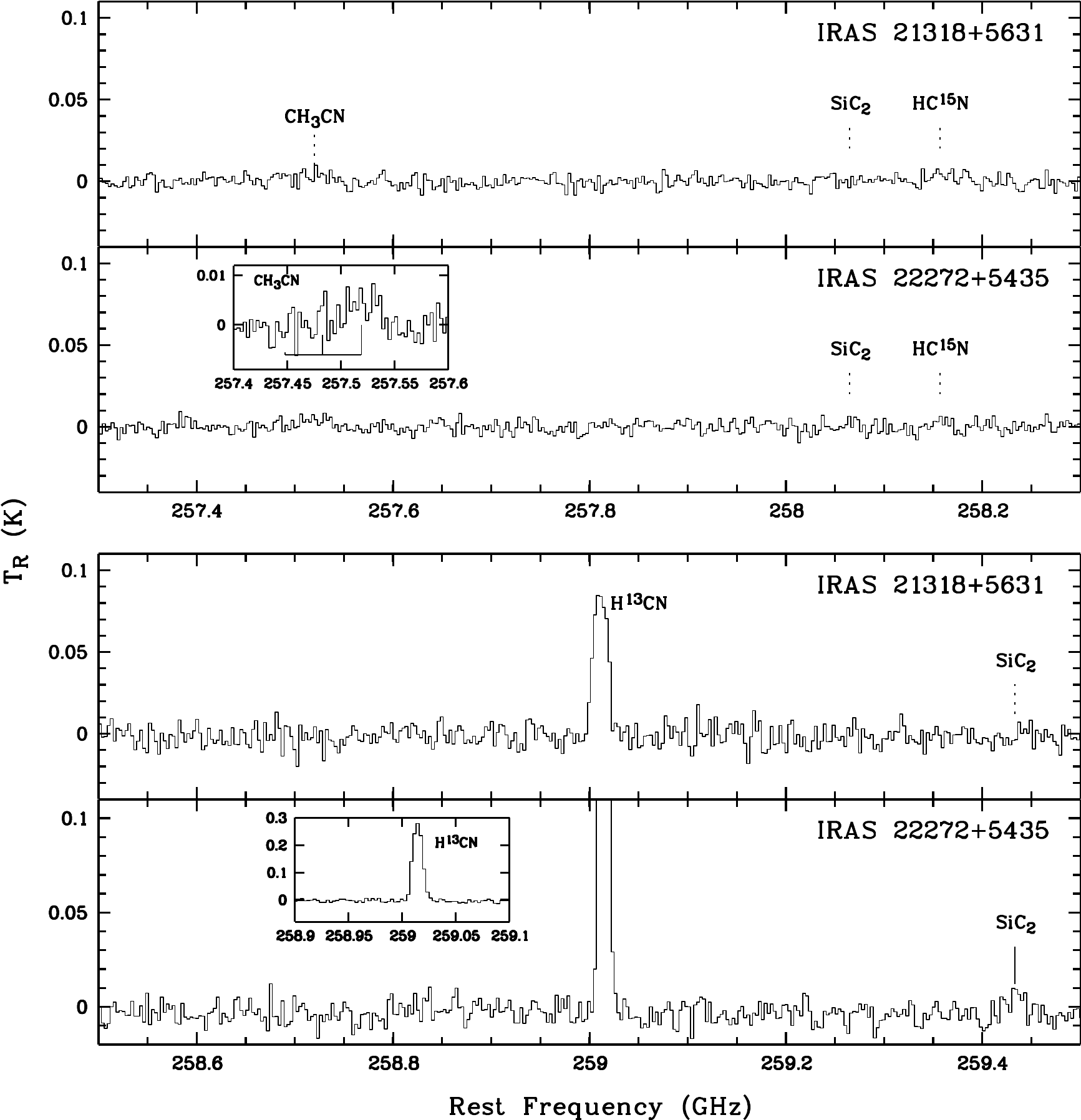,height=16cm}
\end{figure*}

\begin{figure*}
\epsfig{file=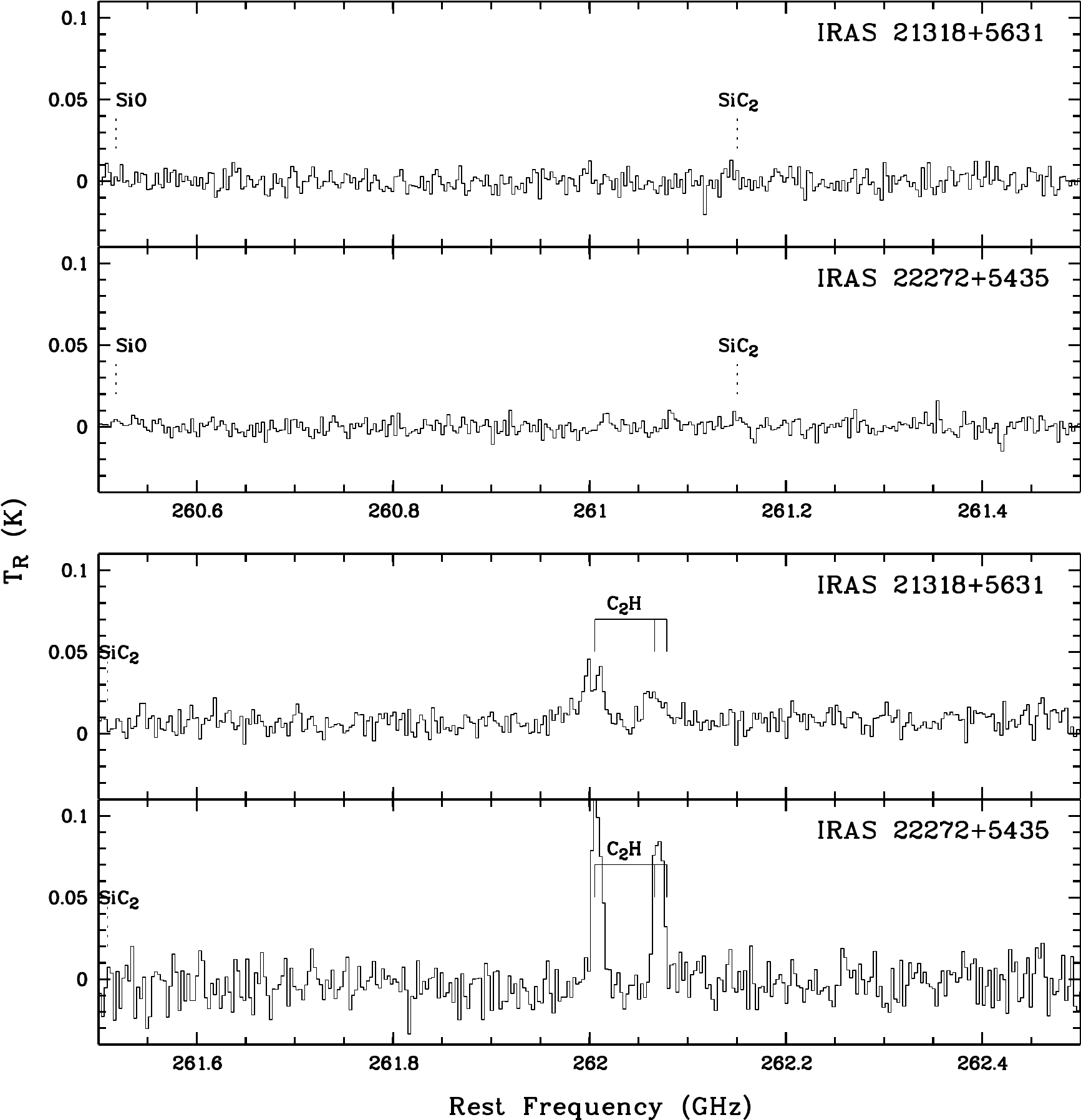,height=16cm}
\end{figure*}

\begin{figure*}
\epsfig{file=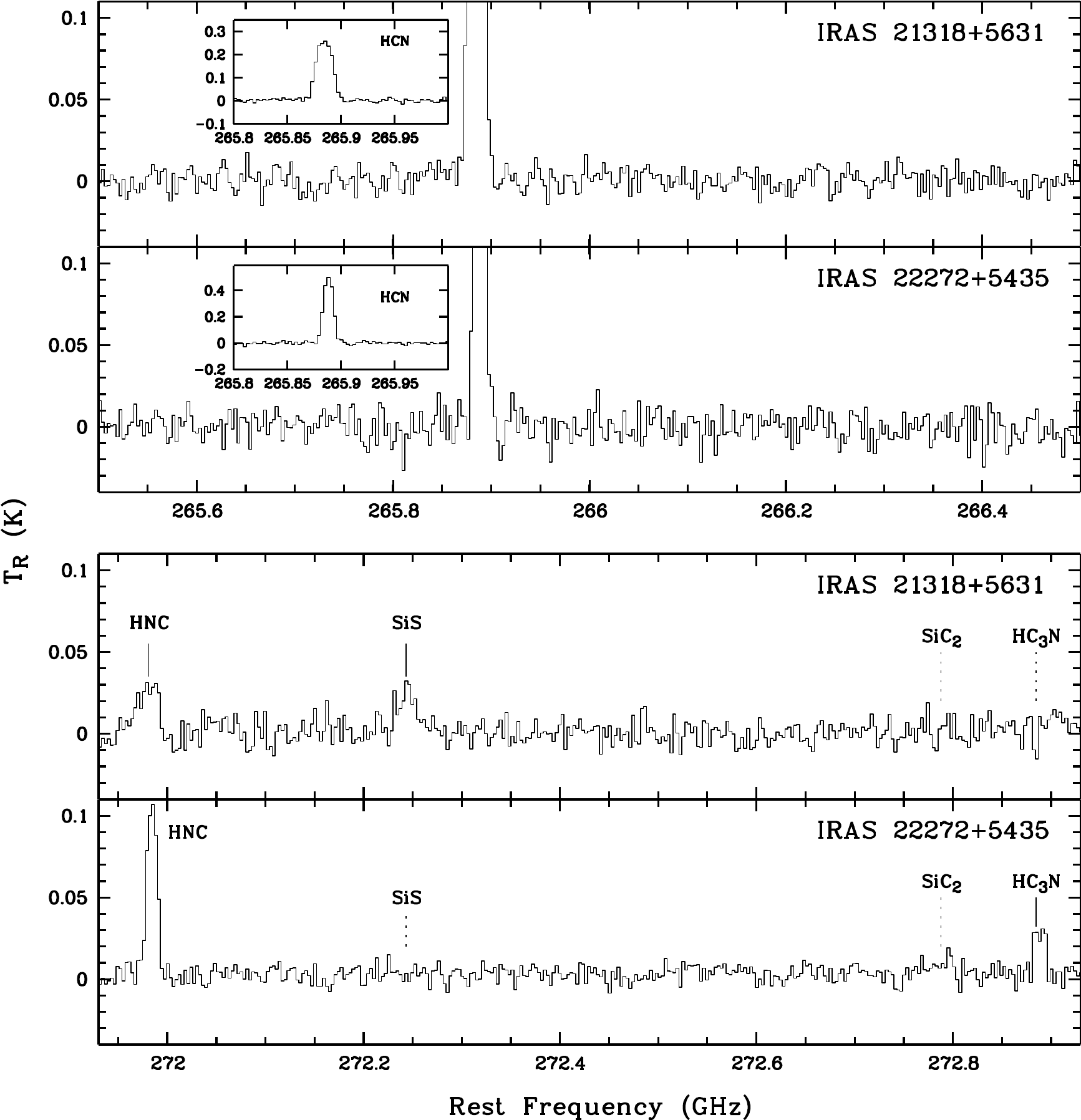,height=16cm}
\end{figure*}

\begin{deluxetable}{lllcrcc@{\extracolsep{0.1in}}crcc}
\rotate
\tablecaption{Molecular transitions detected in IRAS\,21318+5631 and 22272+5435.\label{trans}}
\tabletypesize{\scriptsize}
\tablewidth{0pt}
\tablehead{
\colhead{Frequency}& \colhead{Species} & \colhead{Transition} &  \multicolumn{4}{c}{IRAS\,21318+5631}  & \multicolumn{4}{c}{IRAS\,22272+5435} \\
\cline{4-7}\cline{8-11}
 & & & \colhead{$rms$} &\colhead{$T_R$} &\colhead{$\int T_R\mathrm{d}v$} & \colhead{$\Delta v_{\mathrm{FWHM}}$} & \colhead{$rms$} & \colhead{$T_R$} &\colhead{$\int T_R\mathrm{d}v$} & \colhead{$\Delta v_{\mathrm{FWHM}}$}\\
\colhead{(MHz)} &  & \colhead{(Upper--Lower)} & \colhead{(mK)} & \colhead{(K)} & \colhead{(K\,km/s)} & \colhead{(km/s)} & \colhead{(mK)} & \colhead{(K)} & \colhead{(K\,km/s)} & \colhead{(km/s)}}
\startdata
130268.7 & SiO & $J=3-2$ & 3 & $<$0.009 & $<$0.18 & \nodata & 2 & $<$0.006 & $<$0.06 & \nodata \\
217105.0 & SiO & $J=5-4$ & 4 & $<$0.012 & $<$0.24 & \nodata & 6 & $<$0.006 & $<$0.24 & \nodata \\
260518.1 & SiO & $J=6-5$ & 4 & $<$0.012 & $<$0.24 & \nodata & 3 & $<$0.009 & $<$0.12 & \nodata \\
262004.3     & C$_2$H & $3-2,~J=7/2-5/2,~F=4-3$ & 9 & 0.029 & 0.86$\pm$0.10 & 24.0 & 9 & 0.113 & 1.55$\pm$0.17 & 10.9 \\
262065.0$^a$ & C$_2$H & $3-2,~J=5/2-3/2,~F=3-2$ & 9 & 0.018 & 0.45$\pm$0.07 & 23.6 & 9 & 0.095 & 1.30$\pm$0.19 & 10.9 \\
262067.5$^a$ & C$_2$H & $3-2,~J=5/2-3/2,~F=2-1$ & * & * & * & * & * & * & * & *\\
262078.9$^a$ & C$_2$H & $3-2,~J=5/2-3/2,~F=2-2$ & * & * & * & * & * & * & * & *\\
133213.6 & C$_4$H & $14-13,~J=29/2-27/2$ & 8 & 0.020 & 0.40: & 16: & 7 & 0.030 & 0.30:         & 10: \\
133252.1 & C$_4$H & $14-13,~J=27/2-25/2$ & 8 & 0.025 & 0.50: & 16: & 7 & 0.033 & 0.46$\pm$0.10 & 11.3 \\
142728.8 & C$_4$H & $15-14,~J=31/2-29/2$ & 6 & 0.023 & 0.50$\pm$0.11 & 17.3 & 5 & 0.022 & 0.29$\pm$0.07 & 10.2 \\
142767.3 & C$_4$H & $15-14,~J=29/2-27/2$ & 6 & 0.029 & 0.61$\pm$0.10 & 16.8 & 5 & 0.029 & 0.40$\pm$0.06 & 11.0 \\
152243.6 & C$_4$H & $16-15,~J=33/2-31/2$ & 7 & 0.021 & 0.50$\pm$0.10 & 18.7 & 5 & 0.028 & 0.25$\pm$0.05 & 7.3 \\
152282.1 & C$_4$H & $16-15,~J=31/2-29/2$ & 7 & 0.024 & 0.53$\pm$0.10 & 17.6 & 5 & 0.025 & 0.42$\pm$0.07 & 13.2 \\
161758.1 & C$_4$H & $17-16,~J=35/2-33/2$ & 7 & 0.014 & 0.36$\pm$0.11 & 20.3 & 6 & 0.018 & 0.27$\pm$0.07 & 12.1 \\
161796.6 & C$_4$H & $17-16,~J=33/2-31/2$ & 7 & 0.023 & 0.55$\pm$0.10 & 18.8 & 6 & 0.017 & 0.23$\pm$0.07 & 10.7 \\
228348.6 & C$_4$H & $24-23,~J=49/2-47/2$ & 3 & $<$0.009 & $<$0.15 & \nodata & 3 & $<$0.009 & $<$0.09 & \nodata \\
228387.0 & C$_4$H & $24-23,~J=47/2-45/2$ & 3 & $<$0.009 & $<$0.15 & \nodata & 3 & $<$0.009 & $<$0.09 & \nodata \\
136464.4 & HC$_3$N & $J=15-14$ & 5 & 0.018 & 0.44$\pm$0.08 & 19.8 & 5 & 0.070 & 0.96$\pm$0.05 & 11.0 \\
145561.0 & HC$_3$N & $J=16-15$ & 8 & 0.027 & 0.29$\pm$0.08 & 8: & 9 & 0.065 & 0.97$\pm$0.10 & 11.8 \\
154657.3 & HC$_3$N & $J=17-16$ & 9 & $<$0.017 & $<$0.36 & \nodata & 7 & 0.047 & 0.67$\pm$0.08 & 11.3 \\
163753.4 & HC$_3$N & $J=18-17$ & 6 & $<$0.018 & $<$0.24 & \nodata & 7 & 0.026 & 0.67$\pm$0.19 & 20.8 \\
218324.7 & HC$_3$N & $J=24-23$ & 3 & $<$0.009 & $<$0.12 & \nodata & 3 & 0.038 & 0.64$\pm$0.06 & 13.3 \\
227418.9 & HC$_3$N & $J=25-24$ & 3 & $<$0.009 & $<$0.12 & \nodata & 4 & 0.047 & 0.64$\pm$0.05 & 10.8 \\
236512.8 & HC$_3$N & $J=26-25$ & 3 & $<$0.009 & $<$0.12 & \nodata & 3 & 0.043 & 0.68$\pm$0.04 & 12.8 \\
245606.3 & HC$_3$N & $J=27-26$ & 3 & $<$0.009 & $<$0.12 & \nodata & 2 & 0.042 & 0.66$\pm$0.04 & 12.5 \\
254699.5 & HC$_3$N & $J=28-27$ & 6 & $<$0.018 & $<$0.24 & \nodata & 4 & 0.030 & 0.45$\pm$0.05 & 11.9 \\
272884.7 & HC$_3$N & $J=30-29$ & 6 & $<$0.018 & $<$0.24 & \nodata & 5 & 0.028 & 0.40$\pm$0.06 & 11.4 \\
137180.7 & SiC$_2$ & $6\left(0,6\right)-5\left(0,5\right)$ & 3 & $<$0.009 & $<$0.21 & \nodata & 3 & 0.013 & 0.17$\pm$0.03 & 10.5 \\
140920.2 & SiC$_2$ & $6\left(2,5\right)-5\left(2,4\right)$ & 2 & 0.005 & 0.13: & 20: & 2 & 0.009 & 0.10$\pm$0.02 & 9.2 \\
145325.8 & SiC$_2$ & $6\left(2,4\right)-5\left(2,3\right)$ & 5 & $<$0.015 & $<$0.33 & \nodata & 7 & $<$0.021 & $<$0.27 & \nodata \\
158499.2 & SiC$_2$ & $7\left(0,7\right)-6\left(0,6\right)$ & 2 & 0.006 & 0.06$\pm$0.02 & 8: & 2 & 0.009 & 0.15$\pm$0.02 & 12.8 \\
220773.7 & SiC$_2$ & $10\left(0,10\right)-9\left(0,9\right)$ & 3 & $<$0.009 &$<$0.21 & \nodata & 4 & 0.014 & 0.21$\pm$0.05 & 11.9 \\
222009.4 & SiC$_2$ & $9\left(2,7\right)-8\left(2,6\right)$ & 3 & $<$0.009 & $<$0.21 & \nodata & 4 & 0.015 & 0.30$\pm$0.06 & 15.4 \\
235713.0 & SiC$_2$ & $10\left(6,5\right)-9\left(6,4\right)$ & 3 & $<$0.009 & $<$0.21& \nodata & 4 & 0.013 & 0.27$\pm$0.06 & 16.7 \\
237150.0 & SiC$_2$ & $10\left(4,7\right)-9\left(4,6\right)$ & 3 & $<$0.009 & $<$0.21& \nodata & 3 & 0.010 & 0.18$\pm$0.05 & 13.8 \\
237331.3 & SiC$_2$ & $10\left(4,6\right)-9\left(4,5\right)$ & 3 & $<$0.009 & $<$0.21& \nodata & 3 & 0.011 & 0.13$\pm$0.04 & 10.0 \\
241367.7 & SiC$_2$ & $11\left(0,11\right)-10\left(0,10\right)$& 3 & $<$0.009 & $<$0.21 & \nodata & 3 & 0.012 & 0.19$\pm$0.04 & 12.3 \\
254981.5 & SiC$_2$ & $11\left(2,10\right)-10\left(2,9\right)$ & 6 & $<$0.018 & $<$0.42 & \nodata & 4 & 0.017 & 0.21$\pm$0.04 & 10.0 \\
258065.1 & SiC$_2$ & $11\left(8,3\right)-10\left(8,2\right)$ & 3 & $<$0.009 & $<$0.21 & \nodata & 3 & $<$0.009 & $<$0.09 & \nodata \\
259433.3 & SiC$_2$ & $11\left(6,6\right)-10\left(6,5\right)$ & 5 & $<$0.015 & $<$0.33 & \nodata & 5 & 0.011 & 0.28$\pm$0.07 & 19.6 \\
261150.7 & SiC$_2$ & $11\left(4,8\right)-10\left(4,7\right)$ & 4 & $<$0.012 & $<$0.27 & \nodata & 3 & $<$0.009 & $<$0.09 & \nodata \\
261509.3 & SiC$_2$ & $11\left(4,7\right)-10\left(4,6\right)$ & 9 & $<$0.027 & $<$0.63 & \nodata & 9 & $<$0.027 & $<$0.33 & \nodata \\
272787.8 & SiC$_2$ & $11\left(2,9\right)-10\left(2,8\right)$ & 6 & $<$0.018 & $<$0.42 & \nodata & 5 & $<$0.015 & $<$0.18 & \nodata \\
145227.0 & SiS & $J=8-7$ & 5 & 0.019 & 0.39$\pm$0.07 & 15.9 & 7 & $<$0.021 & $<$0.27 & \nodata \\
163376.8 & SiS & $J=9-8$ & 6 & 0.018 & 0.30$\pm$0.07 & 13.8 & 7 & $<$0.021 & $<$0.27 & \nodata \\
217817.6 & SiS & $J=12-11$ & 3 & 0.022 & 0.43$\pm$0.06 & 15.8 & 3 & $<$0.009 & $<$0.12 & \nodata \\
235961.4 & SiS & $J=13-12$ & 3 & 0.026 & 0.57$\pm$0.06 & 17.5 & 4 & $<$0.012 & $<$0.15 & \nodata \\
254103.2 & SiS & $J=14-13$ & 4 & 0.023 & 0.55$\pm$0.07 & 18.9 & 3 & $<$0.009 & $<$0.12 & \nodata \\
272243.0 & SiS & $J=15-14$ & 6 & 0.026 & 0.63$\pm$0.09 & 19.4 & 5 & $<$0.015 & $<$0.21 & \nodata \\
146969.0 & CS & $J=3-2$ & 6 & 0.066 & 1.29$\pm$0.07 & 15.5 & 8 & 0.107 & 1.40$\pm$0.07 & 10.5 \\
244935.6 & CS & $J=5-4$ & 6 & 0.081 & 1.48$\pm$0.09 & 14.6 & 6 & 0.143 & 1.90$\pm$0.08 & 10.6 \\
241016.1 & C$^{34}$S & $J=5-4$ & 3 & 0.009 & 0.19$\pm$0.04 & 17.9 & 3 & 0.013 & 0.19$\pm$0.04 & 12.0 \\
230538.0 & CO & $J=2-1$ & 4 & 0.819 & 18.30$\pm$0.38 & 21.0 & 4 & 1.441 & 17.48$\pm$0.18 & 9.7 \\
220398.7 & $^{13}$CO & $J=2-1$ & 3 & 0.062 & 1.63$\pm$0.06 & 24.8 & 3 & 0.124 & 1.91$\pm$0.05 & 12.3 \\
226359.9$^b$& CN & $2-1,~J=3/2-3/2$ & 4 & $<$0.012 & $<$0.70 & \nodata & 5 &0.040 & 0.33$\pm$0.05 & \nodata \\
226663.7$^b$& CN & $2-1,~J=3/2-1/2$ & 3 & 0.009 & 0.04:         & \nodata & 4 & 0.072 & 1.01$\pm$0.07& \nodata \\
226887.4$^b$& CN & $2-1,~J=5/2-3/2$ & 3 & 0.020 & 0.80$\pm$0.07 & \nodata & 4 & 0.089 & 0.92$\pm$0.06 & \nodata\\
257522.4$^b$& CH$_3$CN & $14-13$ & 3 & $<$0.009 & $<$0.18  & \nodata & 3 & 0.008 & 0.33: & \nodata \\
265886.4& HCN & $J=3-2$ & 6 & 0.276 & 5.06$\pm$0.10 & 16.2 & 7 & 0.524 & 6.56$\pm$0.09 & 10.0 \\
259011.8& H$^{13}$CN & $J=3-2$ & 5 & 0.093 & 1.87$\pm$0.08 & 16.0 & 5 & 0.294 & 3.84$\pm$0.06 & 10.4 \\
258156.9& HC$^{15}$N & $J=3-2$ & 3 & $<$0.009 & $<$0.18  & \nodata & 3 & $<$0.009 & $<$0.12 & \nodata \\
271981.1 & HNC & $J=3-2$ & 6 & 0.030 & 0.79$\pm$0.11 & 20.9 & 5 & 0.111 & 1.53$\pm$0.05 & 11.0 \\
\enddata
\tablenotetext{a}{Blended features.}
\tablenotetext{b}{Measured for groups of unsolved fine-structure lines.}
\end{deluxetable}

\begin{deluxetable}{lccc@{\extracolsep{0.1in}}ccc}
\tablecaption{Excitation temperatures, column densities, and abundances with respect to H$_2$$^a$.
\label{abu}}
\tabletypesize{\footnotesize}
\tablewidth{0pt}
\tablehead{
\colhead{Species} & \multicolumn{3}{c}{IRAS\,21318+5631}  & \multicolumn{3}{c}{IRAS\,22272+5435} \\
\cline{2-4}\cline{5-7}
& \colhead{$T_{\rm ex}$\,(K)} & \colhead{$N$\,(cm$^{-2}$)}  & 
\colhead{f$_{\rm X}$} &
\colhead{$T_{\rm ex}$\,(K)} & \colhead{$N$\,(cm$^{-2}$)}  & \colhead{f$_{\rm X}$} 
}
\startdata
SiO       & ...$^b$   &$<4\times10^{13}$        &$<8\times10^{-8}$ &    ...$^c$  & $<2\times10^{13}$& $<4\times10^{-8}$   \\
C$_2$H    & ...$^b$   &5.1(0.6)$\times10^{14}$  &9.4(1.1)$\times10^{-7}$ &    ...$^d$  & 9.1(0.1)$\times10^{14}$ &1.5(0.1)$\times10^{-6}$  \\
C$_4$H    & 29(3)     &6.6(1.4)$\times10^{15}$  &1.2(0.3)$\times10^{-5}$ &   24(8)  & 3.6(2.6)$\times10^{15}$    &5.9(4.3)$\times10^{-6}$  \\
HC$_3$N   &  ...$^b$  & 7.5(1.7)$\times10^{13}$ &1.4(0.3)$\times10^{-7}$ &    54(3) &  2.0(0.3)$\times10^{14}$   &3.3(0.5)$\times10^{-7}$   \\
SiC$_2$      &  ...$^b$  & 8.0(3.0)$\times10^{13}$ &1.4(0.6)$\times10^{-7}$ &    84(21)&  2.0(0.5)$\times10^{14}$   &3.3(0.8)$\times10^{-7}$  \\
SiS       &  57(13)& 2.1(0.7)$\times10^{14}$    &3.9(1.3)$\times10^{-7}$ &    ...$^c$  &   $<7\times10^{13}$     &$<2\times10^{-7}$       \\
CS        &  15:   & 2.8(0.3)$\times10^{14}$    &5.2(0.6)$\times10^{-7}$ &    17:      &  3.4(0.1)$\times10^{14}$&5.6(0.2)$\times10^{-7}$   \\
C$^{34}$S &  ...$^e$  & 3.4(0.7)$\times10^{13}$ &6.3(1.3)$\times10^{-8}$ &    ...$^e$  &  3.2(0.7)$\times10^{13}$&5.2(1.1)$\times10^{-8}$   \\
CO        &  ...$^b$  & 5.7(0.1)$\times10^{17}$ &1.1(0.1)$\times10^{-3}$ &    ...$^d$  &  5.2(0.1)$\times10^{17}$&8.5(0.2)$\times10^{-4}$   \\
$^{13}$CO &  ...$^b$  & 5.4(0.2)$\times10^{16}$ &1.0(0.1)$\times10^{-4}$ &    ...$^d$  &  6.1(0.2)$\times10^{16}$&1.0(0.1)$\times10^{-4}$   \\
CN        &  ...$^b$  & 1.1(0.2)$\times10^{15}$ &2.0(0.4)$\times10^{-6}$ &    ...$^d$  &  1.4(0.3)$\times10^{15}$&2.3(0.5)$\times10^{-6}$   \\
CH$_3$CN  &  ...$^b$  & $<7\times10^{13}$       &$<2\times10^{-7}$       &    ...$^d$  &  1.0:$\times10^{14}$    &1.6:$\times10^{-7}$       \\
HCN       &  ...$^b$  & 1.9(0.1)$\times10^{14}$ &3.5(0.2)$\times10^{-7}$ &    ...$^d$  &  2.4(0.1)$\times10^{14}$&3.9(0.2)$\times10^{-7}$   \\
H$^{13}$CN&  ...$^b$  & 7.2(0.3)$\times10^{13}$ &1.3(0.1)$\times10^{-7}$ &    ...$^d$  &  1.4(0.2)$\times10^{14}$&2.3(0.3)$\times10^{-7}$   \\
HC$^{15}$N&  ...$^b$  & $<7\times10^{12}$       &$<2\times10^{-8}$       &    ...$^d$  &  $<5\times10^{12}$      &$<1\times10^{-8}$        \\
HNC       &  ...$^b$  & 2.7(0.4)$\times10^{13}$ &5.0(0.7)$\times10^{-8}$ &    ...$^d$  &  5.1(0.2)$\times10^{13}$&8.4(0.3)$\times10^{-8}$   \\
\enddata
\tablenotetext{{\it a}}{Numbers in brackets represent error values; colons 
indicate uncertain values. For optically thick species, the given $N$ and f$_{\rm X}$ values
represent their lower limits.}
\tablenotetext{{\it b}}{Assuming $T_{\rm ex}=57$\,K.}
\tablenotetext{{\it c}}{Assuming $T_{\rm ex}=84$\,K.}
\tablenotetext{{\it d}}{Assuming $T_{\rm ex}=54$\,K.}
\tablenotetext{{\it e}}{The $T_{\rm ex}$ values derived from CS are
assumed.}
\end{deluxetable}

\begin{deluxetable}{llcccc}
\tablecaption{Isotopic abundance ratios.
\label{isot}}\tabletypesize{\scriptsize}
\tablewidth{0pt}
\tablehead{
\colhead{Isotopic ratio}& \colhead{Species}& \colhead{IRAS\,21318+5631}
& \colhead{IRAS\,22272+5435}   & \colhead{IRC+10216$^a$} & \colhead{Solar$^b$}\\
}
\startdata
$^{12}$C/$^{13}$C  &$^{12}$CO/$^{13}$CO   & $>12$ & $>9$ &   $45\pm3$       & 89   \\
                   &H$^{12}$CN/H$^{13}$CN & $>3$  & $>2$ &    \nodata       &   \nodata  \\
$^{32}$S/$^{34}$S  & C$^{32}$S/C$^{34}$S  & $>8$  & $>11$& $21.8\pm2.6$     & 22.5 \\
$^{14}$N/$^{15}$N  & HC$^{14}$N/HC$^{15}$N& $>27$ & $>39$    & $>4400$      & 272 \\
& H$^{13}$C$^{14}$N/H$^{12}$C$^{15}$N$^c$& $>292$ & $>1035$    & \nodata      & \nodata \\
\enddata
\tablenotetext{a}{From the references in \citet{kah00}.}
\tablenotetext{b}{From \citet{lod03}.}
\tablenotetext{c}{Assume that the $^{12}$C/$^{13}$C ratio is 
the same as that in IRC+10216.}
\end{deluxetable}

\end{document}